\documentstyle[epsfig]{mn}

\newif\ifAMStwofonts

\def \eg{e.g.,\/}
\def \ie{i.e.,\/}
\def \Hb{${\rmn H}\beta$}
\def \hbp{${\rmn H}\beta \, ^\prime$}

\def \Hg{${\rmn H}\gamma$}

\def \mgb{Mg{\,\it b}}
\def \fe{$<$Fe$>$}
\def \mgbp{Mg{\,\it b}\,$^\prime$}
\def \fep{$<\rmn{Fe}\,^\prime\negthinspace>$}
\def \kms {\rm km~s$^{-1}$}

\def \OIIIb {[O{\small III}]$\lambda 5007$}
\def \NI {[N{\small I}]$\lambda 5199$}
\def \gon {Gonz\'{a}lez}

\ifoldfss
  \newcommand{\rmn}[1] {{\rm #1}}

  \ifCUPmtlplainloaded \else
    \NewTextAlphabet{textbfit} {cmbxti10} {}
    \NewTextAlphabet{textbfss} {cmssbx10} {}
    \NewMathAlphabet{mathbfit} {cmbxti10} {} 
    \NewMathAlphabet{mathbfss} {cmssbx10} {} 
  \fi
  \ifAMStwofonts
    \ifCUPmtlplainloaded \else
      \NewSymbolFont{upmath} {eurm10}
      \NewSymbolFont{AMSa} {msam10}
      \NewMathSymbol{\upi}     {0}{upmath}{19}
      \NewMathSymbol{\umu}     {0}{upmath}{16}
      \NewMathSymbol{\upartial}{0}{upmath}{40}
      \NewMathSymbol{\leqslant}{3}{AMSa}{36}
      \NewMathSymbol{\geqslant}{3}{AMSa}{3E}

       \let\le=\leqslant
       \let\ge=\geqslant
    \fi
  \fi
\fi 

\ifnfssone
  \newmathalphabet{\mathit}
  \addtoversion{normal}{\mathit}{cmr}{m}{it}
  \addtoversion{bold}{\mathit}{cmr}{bx}{it}
  \newcommand{\rmn}[1] {\mathrm{#1}}

  \newmathalphabet{\mathbfit} 
  \addtoversion{normal}{\mathbfit}{cmr}{bx}{it}
  \addtoversion{bold}{\mathbfit}{cmr}{bx}{it}
  \newmathalphabet{\mathbfss} 
  \addtoversion{normal}{\mathbfss}{cmss}{bx}{n}
  \addtoversion{bold}{\mathbfss}{cmss}{bx}{n}
  \ifAMStwofonts
    \ifCUPmtlplainloaded \else
      \UseAMStwoboldmath
      \makeatletter
      \new@mathgroup\upmath@group
      \define@mathgroup\mv@normal\upmath@group{eur}{m}{n}
      \define@mathgroup\mv@bold\upmath@group{eur}{b}{n}
      \edef\UPM{\hexnumber\upmath@group}
      \new@mathgroup\amsa@group
      \define@mathgroup\mv@normal\amsa@group{msa}{m}{n}
      \define@mathgroup\mv@bold\amsa@group{msa}{m}{n}
      \edef\AMSa{\hexnumber\amsa@group}
      \makeatother
      \mathchardef\upi="0\UPM19
      \mathchardef\umu="0\UPM16
      \mathchardef\upartial="0\UPM40
      \mathchardef\leqslant="3\AMSa36
      \mathchardef\geqslant="3\AMSa3E

       \let\le=\leqslant
       \let\ge=\geqslant
    \fi
  \fi
\fi 

\ifnfsstwo
  \newcommand{\rmn}[1] {\mathrm{#1}}

  \DeclareMathAlphabet{\mathbfit}{OT1}{cmr}{bx}{it}
  \SetMathAlphabet\mathbfit{bold}{OT1}{cmr}{bx}{it}
  \DeclareMathAlphabet{\mathbfss}{OT1}{cmss}{bx}{n}
  \SetMathAlphabet\mathbfss{bold}{OT1}{cmss}{bx}{n}
  \ifAMStwofonts
    \ifCUPmtlplainloaded \else
      \DeclareSymbolFont{UPM}{U}{eur}{m}{n}
      \SetSymbolFont{UPM}{bold}{U}{eur}{b}{n}
      \DeclareSymbolFont{AMSa}{U}{msa}{m}{n}
      \DeclareMathSymbol{\upi}{0}{UPM}{"19}
      \DeclareMathSymbol{\umu}{0}{UPM}{"16}
      \DeclareMathSymbol{\upartial}{0}{UPM}{"40}
      \DeclareMathSymbol{\leqslant}{3}{AMSa}{"36}
      \DeclareMathSymbol{\geqslant}{3}{AMSa}{"3E}

       \let\le=\leqslant
       \let\ge=\geqslant
    \fi
  \fi
\fi 

\ifCUPmtlplainloaded \else
  \ifAMStwofonts \else 
    \def\upi{\pi}
    \def\umu{\mu}
    \def\upartial{\partial}
  \fi
\fi

\title[Spectroscopic indices of early-type galaxies]{On the dependence
  of spectroscopic indices of early-type galaxies on age, metallicity
  and velocity dispersion}

\author[Harald Kuntschner et~al.] {Harald Kuntschner$^1$,
  John~R.~Lucey$^1$, Russell 
  J. Smith$^{1,2}$, Michael~J.~Hudson$^3$ \cr and Roger~L.~Davies$^1$\\
  $^1$ University of Durham, Department of Physics, 
  South Road, Durham DH1 3LE, UK\\
  $^2$ Departamento de Astronom\'\i a y Astrof\'\i sica, P. Univ. 
  Cat\'olica de Chile, Casilla 306, Santiago 22, Chile (present address)\\
  $^3$ Department of Physics, University of Waterloo, Waterloo ON N2L
  3G1, Canada 
}
\date{submitted ... accepted ...}

\pagerange{\pageref{firstpage}--\pageref{lastpage}}
\pubyear{2000}

\begin{document}

\maketitle

\label{firstpage}

\begin{abstract}
  We investigate the Mg--$\sigma$ and \fe\/--$\sigma$ relations in a
  sample of 72 early-type galaxies drawn mostly from cluster and group
  environments using a homogeneous data-set which is well-calibrated
  onto the Lick/IDS system. The small intrinsic scatter in Mg at a
  given $\sigma$ gives upper limits on the spread in age and
  metallicity of $49$\% and $32$\% respectively, if the spread is
  attributed to one quantity only and if the variations in age and
  metallicity are uncorrelated. The age/metallicity distribution as
  inferred from the \Hb\/ {\em vs}\/ \fe\/ diagnostic diagram
  reinforces this conclusion, as we find mostly galaxies with large
  luminosity weighted ages spanning a range in metallicity.  Using
  Monte-Carlo simulations, we show that the galaxy distribution in the
  \Hb\/ {\em vs}\/ \fe\/ plane cannot be reproduced by a model in which
  galaxy age is the only parameter driving the index-$\sigma$ relation.
  In our sample we do not find significant evidence for an
  anti-correlation of ages and metallicities which would keep the
  index--$\sigma$ relations tight while hiding a large spread in age
  and metallicity. As a result of correlated errors in the
  age-metallicity plane, a mild age-metallicity anti-correlation cannot
  be completely ruled out given the current data. Correcting the
  line-strengths indices for non-solar abundance ratios following the
  recent paper by Trager~et~al., leads to higher mean metallicity and
  slightly younger age estimates while preserving the metallicity
  sequence. The [Mg/Fe] ratio is mildly correlated with the central
  velocity dispersion and ranges from [Mg/Fe]~$=$~0.05 to 0.3 for
  galaxies with $\sigma > 100$~\kms. Under the assumption that there is
  no age gradient along the index--$\sigma$ relations, the
  abundance-ratio corrected Mg--$\sigma$, Fe--$\sigma$ and
  \Hb--$\sigma$ relations give consistent estimates of $\Delta {\rmn
    [M/H]}/ \Delta \log \sigma \simeq 0.9 \pm 0.1$. The slope of the
  \Hb--$\sigma$ relation limits a potential age trend as a function of
  $\sigma$ to 2-3 Gyrs along the sequence.

\end{abstract}

\begin{keywords}
  galaxies: abundances - galaxies: formation - galaxies: elliptical and
  lenticular - galaxies: kinematics and dynamics
\end{keywords}

\section{INTRODUCTION}
\label{sec:intro}
Over the last two decades spectroscopic and photometric observations of
nearby early-type galaxies have shown that they obey tight scaling
relations. For example more luminous galaxies are redder
\cite{san78,lar80,bow92}; the strength of the Mg-absorption (at
$\sim$5175 \AA) increases with increasing central velocity dispersion
\cite{ter81,ben93,coll99}; and early-type galaxies populate only a band
(or `plane') in the three dimensional space of central velocity
dispersion, effective radius, and mean effective surface brightness,
the so called `Fundamental Plane' \cite{djo87,dre87}.

The tightness of the scaling relations has traditionally been
interpreted as evidence for a very homogeneous population of early-type
galaxies, \ie\/ old stellar populations and similar dynamical make-up
of the galaxies \cite{bow92,ren93,ell97,vdo98}. For example the
colour-magnitude relation is perhaps best explained by a correlation of
the metal abundance of the stellar population and total galaxy mass
with no age gradient along the sequence \cite{bow92,kod98,vaz2000}.
However, evidence is being found that elliptical galaxies do generally
have complicated dynamical structures (\eg\/ decoupled cores) and
disturbed morphologies (\eg\/ shells), suggesting an extended and
complex assembly such as in a hierarchical merging picture (\eg\/ Cole
et~al. 2000). Furthermore measurements of absorption line-strength
indices together with evolutionary population synthesis models suggest
that the mean age of the stellar populations in early-type galaxies
span a wide range \cite{gon93,tra98}.

There is a hint of a connection between the detailed morphology and the
mean ages of the stellar populations as ellipticals with disky
isophotes show on average younger ages \cite{dej97,tra97}. This is in
agreement with the studies of cluster galaxies at modest look-back
times of 3-5 Gyr where an increased fraction of blue, late-type
galaxies is found \cite{but78,but84}. Probably this change is
associated with the transformation of spiral galaxies into S0 types
observed over the same interval \cite{dre97,cou98}. How can we
reconcile the existence of tight scaling relations for present day
early-type galaxies with the evidence of an extended galaxy assembly
and star formation history? An answer to this question rests largely on
our understanding of the detailed physical processes which determine
the spectral or photometric properties seen in scaling relations.

Broad band colours and line-strength indices of integrated stellar
populations are sensitive to changes in both their age and metallicity.
In fact it is possible to find many different combinations of ages and
metallicities which produce similar overall photometric properties
(age-metallicity degeneracy, Worthey 1994). In order to explain tight
scaling relations, despite the presence of a substantial diversity in
the integrated stellar populations, it has been suggested that the ages
and metallicities of integrated stellar populations `conspire' such
that deviations from a given scaling relation due to age variations are
balanced by an appropriate change in metallicity and vice versa; \eg\/
an age-metallicity anti-correlation \cite{tra98,fer99,tra2000b}. For
example, a re-analysis of the \gon\/ sample by Trager et~al. appears to
show that galaxies with a young stellar component are also more metal
rich. Recently, evidence for a negative age-metallicity correlation has
also been found in other data sets \cite{jor99} and in literature
compilations \cite{ter2000}. In contrast to this Kuntschner \& Davies
(1998) and Kuntschner (2000) did not find strong evidence for the
existence of an age-metallicity anti-correlation in their analysis of
the Fornax cluster.  However, to date there is a lack of large, high
quality samples with which it would be possible to probe the population
of early-type galaxies as a whole.

In this paper we seek to further our understanding of scaling relations
such as Mg--$\sigma$. We investigate the spread in the mean ages and
metallicities {\em at a given mass}\/ but also probe the possibility of
age and metallicity trends {\em along}\/ the relations. We use
established methods such as the analysis of the scatter \cite{coll99}
in the Mg--$\sigma$ relation and line-strength age/metallicity
diagnostic diagrams, and take advantage of recent advances in stellar
population modeling such as improved corrections for non-solar
abundance ratios \cite{tra2000a}.  We employ a homogeneous and
high-quality subset of the data collected initially for the SMAC
peculiar motion survey \cite{hud99,smi2000}. The sample is not complete
but includes galaxies drawn mostly from nearby clusters and groups with
some galaxies from less dense environments.

The present paper is organized as follows. Section~\ref{sec:data}
describes the sample selection, observations, data reduction and the
measurements of the line-strength indices. Section~\ref{sec:mg_sig}
presents the Mg--$\sigma$ and Fe--$\sigma$ relations with an initial
investigation into the scatter and slopes. In Section~\ref{sec:age} we
present a detailed analysis of age-metallicity diagnostic diagrams
including an improved treatment of non-solar abundance ratios. The
effects of the non-solar abundance ratio corrections on the
age-metallicity estimates and on the index--$\sigma$ relations are
presented in Sections~\ref{sec:imp_index} and \ref{sec:imp_sig}
respectively. Our conclusions are given in
Section~\ref{sec:conclusions}.

\section{OBSERVATIONS} 
\label{sec:data}
\subsection{Sample Description}
\label{sec:sample}

The data reported here is from the SMAC I97MA spectroscopic run (see
Smith et~al. 2000 for further details). This run was undertaken
principally to provide a secure connection of the SMAC data to other
data sets and concentrated on observing galaxies with measurements
previously reported by Dressler (1984), \gon\/ (1993) and Davies et~al.
(1987). Accordingly the SMAC I97MA data shows a good overlap with the
Lick/IDS system with 48 galaxies in common. In selecting targets,
higher priority was given to galaxies with measurements from multiple
sources. While the sample is dominated by ellipticals in the Virgo and
Coma clusters, it does include a few S0s and galaxies in less dense
environments (see Table~\ref{tab:table} of the Appendix). It is not a
complete sample in any sense and our sensitivity to environmental
effects is limited. However, we did not find strong differences between
galaxies from different environments or morphologies in our sample and
thus throughout this paper we do not split up our sample but treat it
as a whole.

\subsection{Observations and Data Reduction}
The observations were carried out by JRL at the 2.5m Isaac Newton
Telescope \footnote{ The INT is operated on the island of La Palma by
  the Isaac Newton Group in the Spanish Observatorio del Roque de los
  Muchachos of the Instituto de Astrofisica de Canarias.}  (March
1997). The Intermediate Dispersion Spectrograph was used in conjunction
with the 23.5cm camera, the 900V grating and a Tek1K chip.  With a slit
width of 3~arcsec, an instrumental resolution of $\sim4$~\AA\/ FWHM was
achieved, equivalent to an instrumental dispersion ($\sigma_{\rm
  instr}$) of 98~\kms. The spectra cover a wavelength range of
1024~\AA, sampled at 1~\AA\/ pix$^{-1}$ from 4800 to 5824~\AA\/. In
total 200 galaxy spectra were obtained including many repeat
observations. Along the slit the central five pixels of each
observation were combined giving an effective aperture of
$3.0\times3.4$~arcsec$^2$. For other basic data reduction details and
the velocity dispersion measurements see Smith~et~al. (2000).

The wavelength range and the redshift distribution of our data allows
the measurement of five important line-strength indices (\Hb, Mg$_2$,
\mgb, Fe5270, Fe5335) in the Lick/IDS system (Worthey 1994, hereafter
W94, Trager et~al. 1998). The indices were measured on the galaxy
spectra after they had been corrected to a relative flux scale and were
broadened to the Lick/IDS resolution\footnote{We note, that our
  approach here is different to that of Smith et~al., whose tabulated
  Mg$_2$ and \mgb\/ were measured from the un-broadened spectra. } (see
also Worthey \& Ottaviani 1997). The index measurements were then
corrected for internal velocity broadening of the galaxies following
the method outlined in Kuntschner (2000). The formal measurement errors
were rescaled to ensure agreement with the error estimates derived from
the repeat observations (the scale factors ranged from 0.80 to 1.07 for
all indices but Mg$_2$ with a scale factor of 0.44). In order to
maximize the S/N, multiple index measurements of the same galaxy were
averaged giving a sample of 140 different galaxies.  The effective S/N
per~\AA\/ of these combined measurements ranges from 14 to 78. In order
to keep the errors of the index measurements reasonably low we decided
to use only index measurements with an effective $\rmn{S/N} \ge 30$
per~\AA\/ giving a final sample of 72 galaxies (median S/N~$\sim 40$
per~\AA) which we will analyze in this paper.  \footnote{N4278 was also
  excluded from the final sample because it shows strong \Hb\/,
  \OIIIb\/ and \NI\/ emission which severely affect the \Hb\/ index and
  \mgb\/ index.}

All of our data were obtained using the same physical aperture size,
but since the galaxies span a factor of $\sim$10 in distance, the
spectra sample different physical aperture sizes in kpc. Since line
indices and velocity dispersion exhibit measurable gradients with
respect to radius, it is necessary to correct these parameters to a
`standard' aperture. Here we adopt a generalisation of the aperture
correction due to J{\o}rgensen, Franx \& Kj{\ae}rgaard (1995), which
scales parameters to a circular aperture of diameter 1.19 $h^{-1}$~kpc,
$H_0 = 100~h~$\kms\/ Mpc$^{-1}$, equivalent to a circular aperture of
3.4 arcsec at the distance of Coma. The strength of the correction
(\ie\/ the correction in the quantity per dex in aperture size) is
$-0.04$ for $\log \sigma$, Mg$_2$ and \mgb, but $-0.02$ for \fe\/ and
zero (\ie\/ no correction) for \Hb. These correction strengths are
based on the average line strength gradients observed in early-type
galaxies \cite{kun98b}. The correction strengths adopted in this paper
are overall similar to the ones used in J{\o}rgensen (1997), however we
use $-0.02$ for the \fe\/ index, whereas J{\o}rgensen used $-0.05$.

\subsection{Matching the Lick/IDS system}
In order to determine the systematic offsets between our line-strength
system and the original Lick/IDS system (on which the Worthey 1994
model predictions are based), we compared index measurements for the 48
galaxies in common with the Lick galaxy library \cite{tra98}. We note
that the Lick/IDS aperture was $1.4\times4$~arcsec; so for the
following comparison our data was corrected to the (fixed) Lick/IDS
aperture using the formula given in J{\o}rgensen et~al. (1995).

Generally there is good agreement between the two data sets and only
small offsets are found (see Table~\ref{tab:lick_off} and
Figure~\ref{fig:offset_all}). In order to match the Lick/IDS system we
removed the offsets from our data. The median measurement errors in the
indices, for our sample galaxies, are given in column 4 of
Table~\ref{tab:lick_off}. For a definition of non-standard
line-strength indices see Section~\ref{sec:special_indices}.

\begin{table}
  \caption[]{Lick/IDS offsets and median index errors}
  \label{tab:lick_off}
  \begin{center}

  \begin{tabular}{lclll} \hline 
     Index      & \multicolumn{2}{c}{(Lick/IDS -- this data)} & {index error} & units    \\ \hline 
     H$\beta$   &  $-0.11\pm0.04$   &(0.28)                  &    0.17       & \AA     \\
     Mg$_2$     & $ 0.007\pm0.002$  &(0.014)                 &    0.005      & mag    \\
     \mgb       &  $-0.06\pm0.05$   &(0.36)                  &    0.18       & \AA     \\  
     \fe        &  $-0.08\pm0.04$   &(0.29)                  &    0.16       & \AA     \\ 
     \mgbp      & --                &                        &    0.007      & mag    \\ 
     \fep       & --                &                        &    0.004      & mag    \\ 
     \OIIIb     & --                &                        &    0.10       & \AA     \\ 
     $\sigma_0$ & --                &                        &    0.014      & dex    \\ \hline
    \end{tabular}

    \medskip
    \begin{minipage}{8.0cm}
      Note: There are 48 galaxies in common between the Lick/IDS sample
      and our data set. The quoted offset errors (column 2) reflect the
      error on the mean offset. The rms scatter is given in brackets.
      Column~4 lists the median index error for all indices used in
      this paper.
    \end{minipage}

  \end{center}
\end{table}

\subsection{Comparison with other studies}
We can compare our Lick/IDS-corrected data with the galaxies in common
with \gon\/ (1993), J{\o}rgensen (1999) and Mehlert~et~al.  (2000).
Note, that the former authors also corrected their indices onto the
Lick/IDS system. Any aperture differences were corrected following
J{\o}rgensen et~al. (1995) and J{\o}rgensen (1997).

\begin{figure*}
\epsfig{file=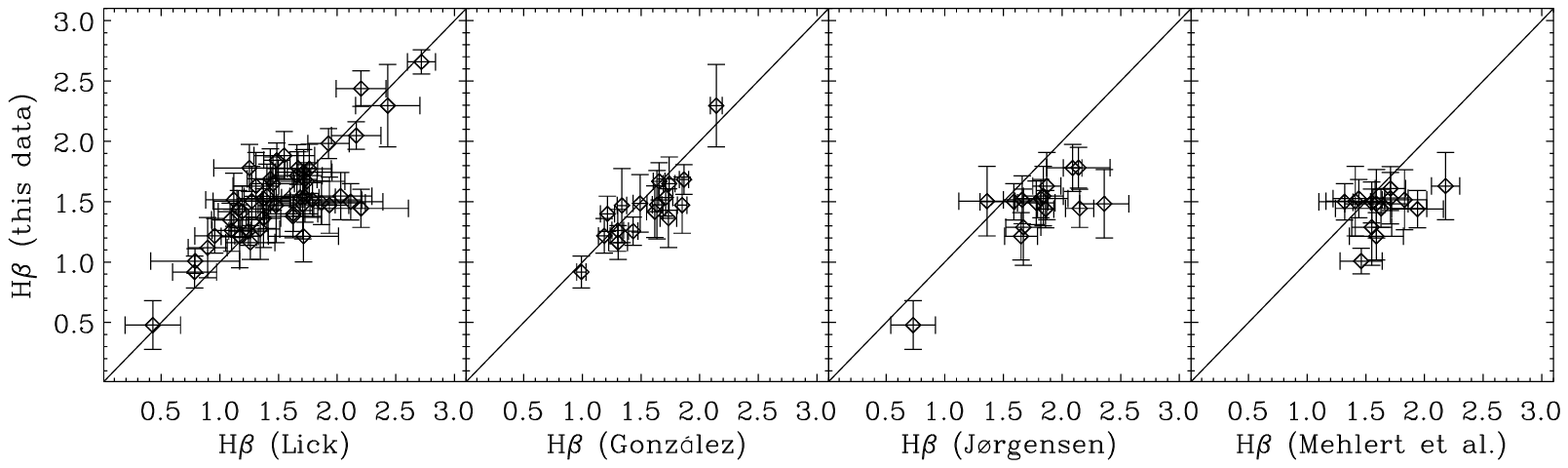,width=15.5cm}
\epsfig{file=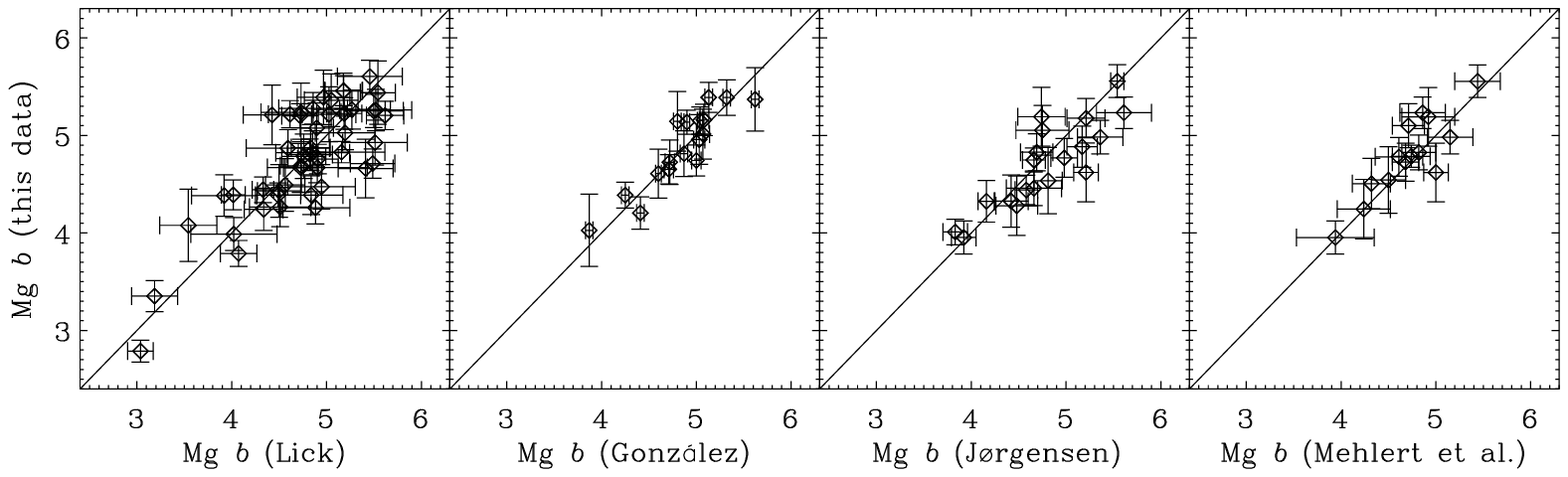,width=15.5cm}
\epsfig{file=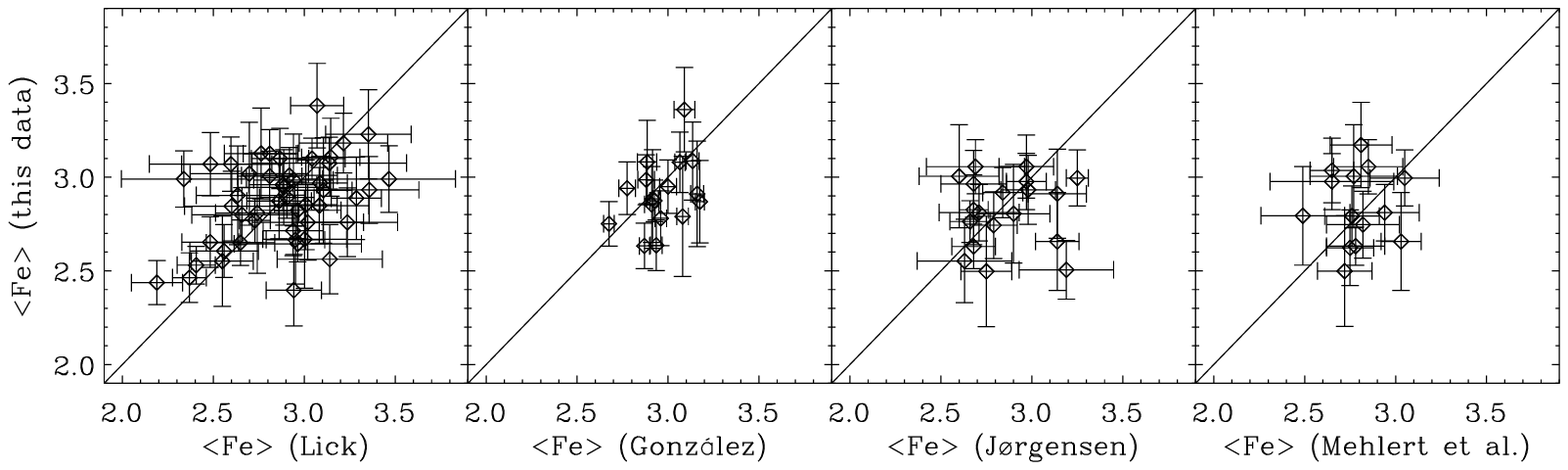,width=15.5cm}
\caption[]{\label{fig:offset_all}Comparisons of our line-strength
  measurements with the literature (Lick data; Trager et~al. 1998,
  \gon\/ 1993, J{\o}rgensen 1999 and Mehlert et~al. 2000) for the \mgb,
  \fe\/ and \Hb\/ indices. Our data was corrected for the Lick/IDS
  offset prior to comparison. The offsets to the data sets of
  \gon\/(1993), J{\o}rgensen (1999) and Mehlert et~al. (2000) are
  summarized in Table~\ref{tab:lit_off}. All indices are measured in
  \AA.}
\end{figure*}

\begin{figure}
\epsfig{file=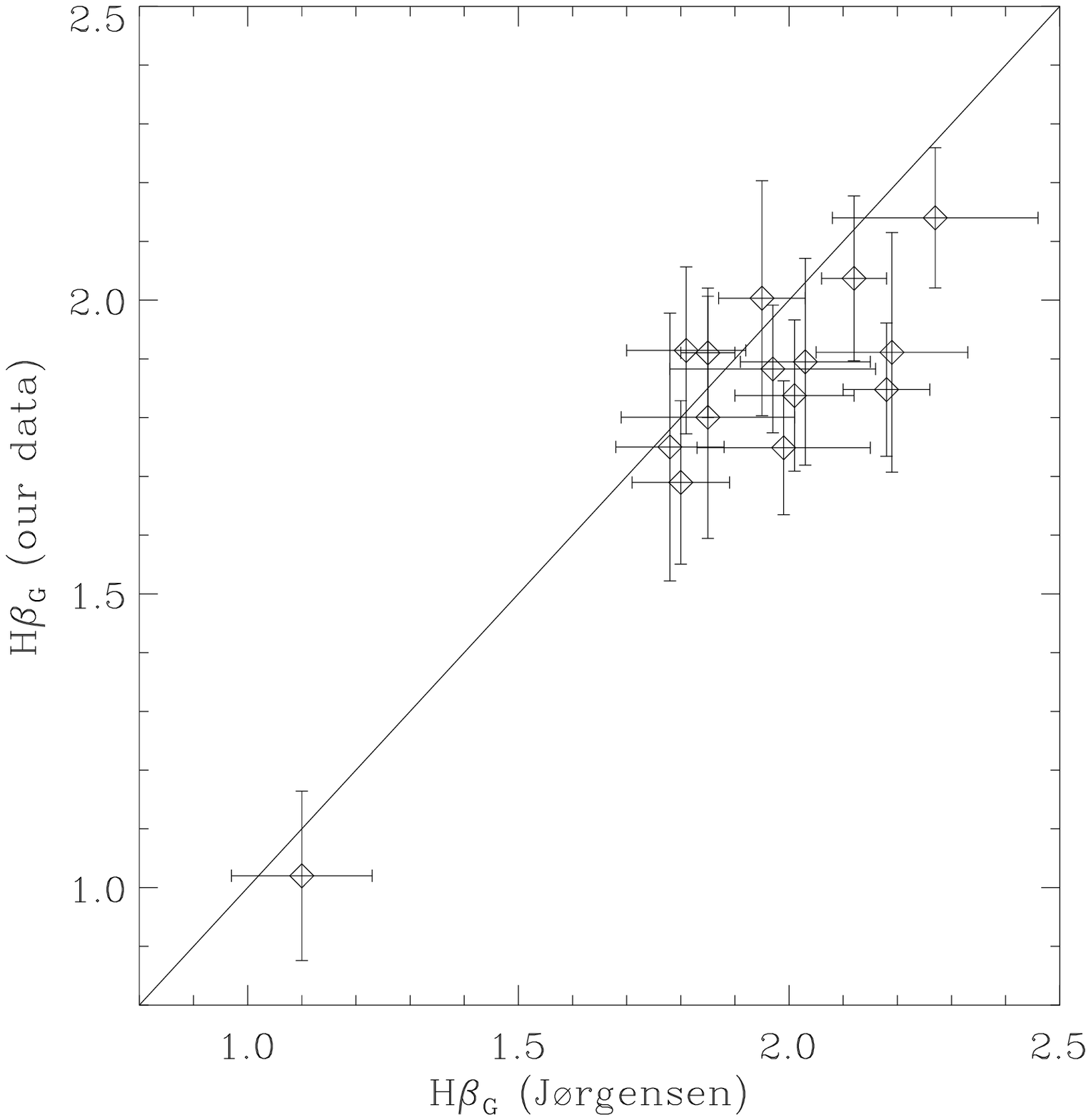,width=8cm} 
\caption[]{\label{fig:hb_G_comp} Comparisons of our line-strength
  measurements of H$\beta_G$ with the 15 galaxies in common with
  J{\o}rgensen (1999). The mean offset is 0.1~\AA\/ (J{\o}rgensen --
  this data) with a rms scatter of 0.12~\AA.}
\end{figure}

For the \mgb\/ and \fe\/ indices there is good general agreement
between our data and the literature (see Table~\ref{tab:lit_off} and
Figure~\ref{fig:offset_all}) with offsets $\le 0.1$~\AA. We also find
good agreement between our data and \gon\/ for the \Hb\/ index. The
Mehlert et~al. data compares less favourably with an offset of
0.19~\AA, while the comparison with J{\o}rgensen shows a large offset
of 0.33~\AA. The offset with respect to J{\o}rgensen is difficult to
interpret, since her data was also calibrated to the Lick/IDS system.
While J{\o}rgensen's spectra were obtained with two different
instruments (LCS and FMOS), both data sets show the same offset. The
data points plotted in Figure~\ref{fig:offset_all} show the combined
data sets of J{\o}rgensen. She also measured a variant of the \Hb\/
index, the so-called H$\beta_G$ index (defined in J{\o}rgensen 1997).
This index is not part of the original Lick/IDS system but has the
advantage of smaller errors compared to \Hb\/ at the same S/N. A
comparison for this index shows a much better agreement between our
data and J{\o}rgensen's (see Figure~\ref{fig:hb_G_comp}) with an offset
of only 0.1~\AA\/ and a small rms scatter of 0.12~\AA. While we do not
have an explanation for the disagreement with J{\o}rgensen, we conclude
that our index measurements are well calibrated onto the Lick/IDS
system with a systematic error $\le 0.1$~\AA.

\begin{table*}
  \caption[]{Offsets to literature data}
  \label{tab:lit_off}
  \begin{center}
   
  \begin{tabular}{lcccc} \hline 
     Index    & Lick -- this data  &\gon\/ -- this data&J{\o}rgensen -- this data&Mehlert et~al. -- this data\\ \hline 
     H$\beta$ & 0.00 $\pm$0.04 (0.28) \AA& +0.09 $\pm$0.04 (0.16) \AA& +0.33 $\pm$0.06 (0.24) \AA& +0.19 $\pm$0.06 (0.23) \AA  \\
     H$\beta_G$&       --                &         --                & +0.10 $\pm$0.03 (0.12) \AA&       --             \\ 
     \mgb     & 0.00 $\pm$0.05 (0.36) \AA&--0.03 $\pm$0.04 (0.17) \AA& +0.07 $\pm$0.06 (0.26) \AA&--0.08 $\pm$0.06 (0.20) \AA  \\  
     $<$Fe$>$ & 0.00 $\pm$0.04 (0.29) \AA& +0.06 $\pm$0.05 (0.18) \AA& +0.03 $\pm$0.06 (0.27) \AA&--0.05 $\pm$0.07 (0.25) \AA  \\
              &                          &                           &                           &                             \\       
\# of galaxies& 48                       & 17                    &  19 (15 for H$\beta$ \& H$\beta_G$)   & 14            \\ \hline
    \end{tabular}

    \medskip
    \begin{minipage}{15cm}
      Note: For the comparison shown here with the Lick data (Trager et
      al. 1998), \gon\/ (1993), J{\o}rgensen (1999) and Mehlert et~al.
      (2000) our data was corrected for the Lick/IDS offset as
      summarized in Table~\ref{tab:lick_off}. Errors on the mean offset
      are given. The rms scatter is given in brackets. The number of
      galaxies in common with each dataset from the literature is given
      in the last row.
    \end{minipage}

  \end{center}
\end{table*}

The Smith et~al. velocity dispersion measurements have been transformed
onto the SMAC standard system by adding the correction
$\Delta\log\sigma=-0.0084\pm0.0046$, determined by Hudson et~al. (2000)
from simultaneous comparisons of many data sources. This correction is
negligibly small for the purposes of this paper.

Our fully corrected line-strength and central velocity dispersion
measurements used in this study are summarized in Table~\ref{tab:table}
of the Appendix.

\subsection{Special indices}
\label{sec:special_indices}
For the line-strength analysis in this paper we will use a Fe indicator
which is the mean of the Lick Fe5270 and Fe5335 indices in order to
suppress measurement errors \cite{gor90}:

\begin{equation}
\label{equ:fe}
  <{\rmn Fe}> = ({\rmn Fe5270} + {\rmn Fe5335}) / 2
\end{equation}

Furthermore, following the papers by Colless~et~al. (1999),
Smith~et~al. (2000) and Kuntschner (2000) we will sometimes express the
line-strengths in magnitudes of absorbed flux, indicating this usage by
a prime after the index (\eg\/ \mgbp)

\begin{equation}
  \label{equ:mgbp}
  {\rmn index}\,^\prime= -2.5 \log \left( 1 - \frac{\rmn index}{\Delta \lambda}\right)  
\end{equation} 
where $\Delta \lambda$ is the width of the index bandpass. For \fe\/ we
define \fep~=~(Fe5270\,$^\prime$ + Fe5335\,$^\prime$)/2. Note, that
the Mg$_2$ index is always expressed in magnitudes.

\begin{figure}
\epsfig{file=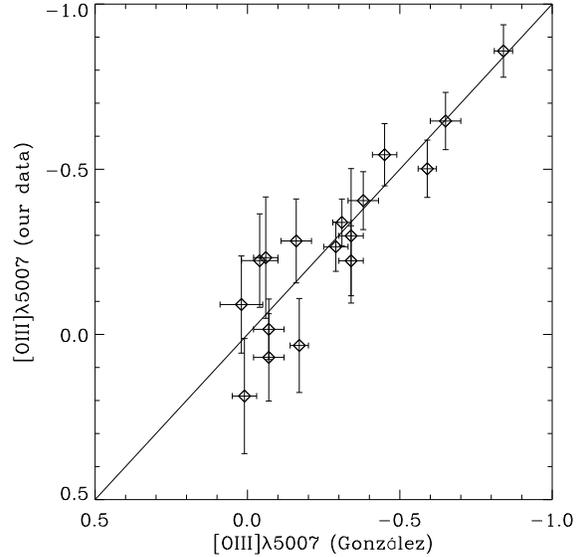,width=8cm} 
\caption[]{\label{fig:emi_gon}Comparison of \OIIIb\/ emission estimates 
  between our data and the galaxies in common with \gon . Both indices
  are measured in \AA. The solid line is the line of equality and not
  fitted to the data. See text for details.}
\end{figure}

In order to estimate the amount of nebular emission present in
galaxies, we defined a \OIIIb\/ index similar to the other Lick/IDS
indices, with continuum passbands of 4978--4998~\AA\/ and
5015--5030~\AA\/ and a central passband of 4998--5015~\AA. Our
measurements of the \OIIIb\/ emission were then compared with the
\gon\/ (1993) measurements for the 17 galaxies in common. The agreement
between our measurements and \gon's is excellent (see
Figure~\ref{fig:emi_gon}). We note, that \gon\/ measured the \OIIIb\/
emission after subtracting the spectrum of the host galaxy whereas our
\OIIIb\/ index was measured directly on the spectra, leading to an
effective absorption measurement when no emission is present. In order
to transform our \OIIIb\/ index measurements into estimates of the true
\OIIIb\/ emission we subtracted an offset of 0.41~\AA\/ (the values
listed in Table~\ref{tab:table} of the Appendix reflect our true
emission estimates).

Forty-three galaxies in our sample show significant (1$\sigma$ level)
\OIIIb\/ emission with a median value of 0.27~\AA. Such emission is
indicative of emission in the Balmer lines which will fill in the {\em
  stellar}\/ \Hb-absorption and hamper accurate age estimates.  \gon\/
(1993) corrected his \Hb\/ data for emission fill-in by adding $0.7
\times$ the \OIIIb\/ emission to \Hb. Trager~et~al. (2000a) revisited
this issue and concluded that the factor should be reduced to 0.6 while
confirming the validity of the procedure. Although it is doubtful
whether this correction is accurate for an individual galaxy (see
Mehlert et~al.  2000), we believe it is a good correction in a
statistical sense.  Therefore we correct our \Hb\/ measurements by
adding $0.6 \times$ the \OIIIb\/ emission.

\section{The \mgbp\/--$\sigma$ and \fep\/--$\sigma$ relations}
\label{sec:mg_sig}
In this section we investigate the scatter in the Mg--$\sigma$ relation
and also compare the slopes of the Mg- and Fe--$\sigma$ relations with
predictions from stellar population models.

\begin{figure}
\epsfig{file=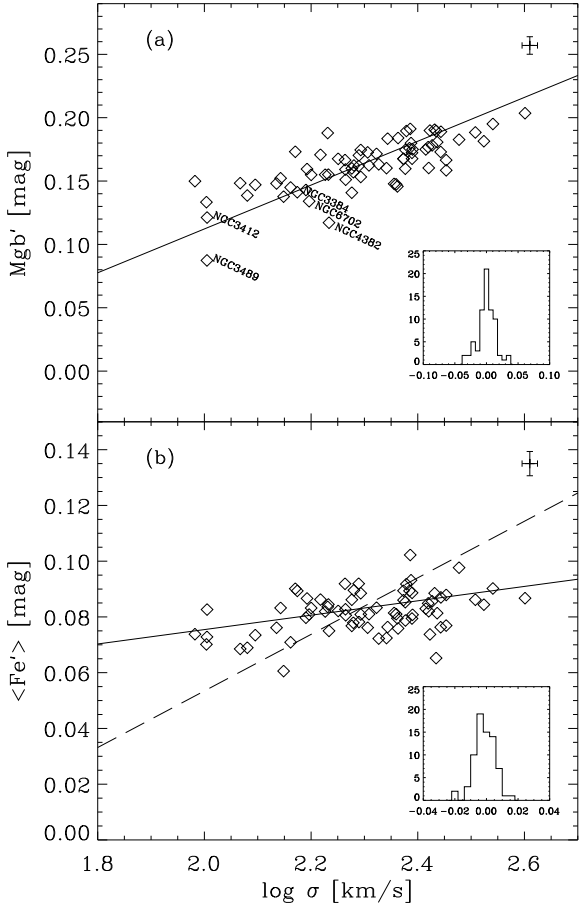,width=8.4cm}
\caption[]{\label{fig:mg_fe_sig1}(a) The \mgbp\/--$\sigma$ relation and  
  (b) the \fep\/--$\sigma$ relation for our sample of 72 early-type
  galaxies. The solid lines indicate a straight line fit taking into
  account errors in x- and y-direction. Error bars representing the
  median errors in each quantity are shown in the upper right corner of
  the panels. The distribution of residuals in \mgb\/ and \fe\/ about
  the fits are shown in the insets. The galaxies with attached names
  are identified in Section~\ref{sec:age} as having younger stellar
  populations. The long-dashed line in panel (b) shows the expected
  slope of the \fep\/--$\sigma$ relation when the metallicity change of
  the \mgbp\/--$\sigma$ relation is assumed. See text for details.}
\end{figure}

Figure~\ref{fig:mg_fe_sig1}a shows the \mgbp\/--$\sigma$ relation for
our sample of 72 galaxies. A straight line fit taking into account the
errors in x- and y-direction (IDL implementation of {\tt fitexy}
routine in Numerical Recipes, Press et~al. 1992) gives:

\begin{equation}
  \label{eq:metal_sig2}
  {\rmn Mg{\,\it b}}\,^\prime = 0.142(\pm0.013) \log \sigma - 0.163(\pm0.031) \, .
\end{equation}

The observed scatter (1$\sigma$ y-residuals) about the best fitting
\mgbp\/--$\sigma$ relation is of 0.0137~mag. Given a median
observational error of 0.007~mag in \mgbp\/ we estimate the intrinsic
scatter by requiring:

\begin{equation}
  \label{eq:intrinsic}
  \sum_{k=1}^{72} \frac{\Delta {\rmn Mg}\, b_k^{\prime 2}}{\delta {\rmn Mg}\, b_k^{\prime 2} + \delta i^2} \equiv 1
\end{equation}
where $\Delta {\rmn Mg}\, b_k^\prime$ is the deviation from the best
fitting relation for each galaxy, $\delta {\rmn Mg}\, b_k^\prime$ are
the individual errors in \mgbp\/ and $\delta i$ is the estimated
intrinsic scatter. This estimator gives an {\em intrinsic}\/ scatter of
0.0117~mag in \mgbp.

The slope and zero point of the relation are in good agreement with
Colless~et~al. (1999) who found \mgbp~$= (0.131\pm0.017) \log \sigma -
(0.131\pm0.041)$ for their EFAR sample of 714 early-type galaxies in
clusters. However, for their sample they find an intrinsic scatter of
0.016~mag in \mgbp\/ which is larger than in our data set.

With the help of stellar population models one can translate the
intrinsic spread in \mgbp\/, at a given $\sigma$, into age and
metallicity spreads. To simplify this exercise we assume initially that
there are no other sources of scatter and that age and metallicity are
not correlated, or only mildly so (see Colless~et~al. 1999 for a more
detailed analysis). Using the Colless~et~al. (1999) calibration of
\mgbp\/ as a function of $\log$~age and metallicity we find for our
data set at a given $\sigma$ and at fixed metallicity a spread of
$\delta t/t = 49$\% in age and at fixed age a spread of $\delta Z/Z =
32$\% in metallicity. Colless~et~al. found 67\% and 43\% respectively.

Our estimates of the spread in age and metallicity can be taken as
upper limits because other effects such as aperture differences and
varying abundance ratios will be responsible for some of the scatter.
Recalling that our sample is a collection from several clusters, groups
and the field, the small scatter around the Mg--$\sigma$ relation is
indeed remarkable and could be interpreted as evidence for homogeneous
stellar populations in early-type galaxies, following a trend of
increasing metallicity with increasing central velocity dispersion or
luminosity.

The prominence of the Mg-$\sigma$ relation would suggest that other
metal line strength indices should also exhibit a correlation with the
central velocity dispersion. Until recently \cite{kun2000} it was
unclear whether a {\em significant}\/ relation exists
\cite{fis96,jor97,jor99}. In Figure~\ref{fig:mg_fe_sig1}b we show the
\fep\/--$\sigma$ relation for the galaxies in our sample. A
(non-parametric) Spearman rank-order test shows a weak correlation
coefficient of 0.34 with a significance level of $0.3$\%. The best
fitting relation is

\begin{equation}
  \label{eq:metal_sig3}
  <{\rmn Fe\,^\prime}> = (0.021\pm0.006) \log \sigma + (0.034\pm0.015) \, .
\end{equation}

Consistent with results for the Fornax cluster \cite{kun2000}, we find
a weak correlation between the \fep\/ index and $\sigma$ for our
sample. In agreement with the findings of J{\o}rgensen (1997,1999) the
relation is very shallow compared to the \mgbp\/--$\sigma$ relation.
Therefore any precise determination of the slope is hampered by its
sensitivity to the size of the observational errors and selection
effects; in particular the sampling of the low velocity dispersion
range is crucial. Recent work by Concannon, Rose \& Caldwell (2000)
indicates that the spread in age at a given $\sigma$ increases towards
the low velocity dispersion range. Hence incompleteness at low velocity
dispersions is potentially a source for bias. Nevertheless, most of the
currently available samples (including our sample) are not complete,
often lacking low luminosity galaxies.

Surprisingly, however, the \fep\/--$\sigma$ relation is not as steep as
one would expect if the \mgb\/ and \fe\/ indices are measures of the
same quantity, \ie\/ metallicity (see also J{\o}rgensen 1997,
Kuntschner 1998, J{\o}rgensen 1999). This can be demonstrated by the
following exercise: Assuming that there is no significant age trend
along the index--$\sigma$ relation one can use stellar population
models to translate the change in index strength per $\log \sigma$ into
a change of metallicity. The expected changes in \mgbp\/ and \fep\/ per
dex in metallicity are given in Table~\ref{tab:indexchange}.

\begin{table}
  \caption[]{Predictions for index changes with metallicity}
  \label{tab:indexchange}
  \begin{center}

  \begin{tabular}{lccc} \hline 
                     & \mgbp & \fep  & \hbp  \\ 
     $\Delta$ index  & 0.091 & 0.053 &-0.031 \\ \hline 
    \end{tabular}

    \medskip
    \begin{minipage}{8cm}
      Note: Model predictions for the average index change per dex in
      metallicity. The predictions are based on Worthey (1994) models
      for a fixed age of 12 Gyr and $-0.25 \le \rmn{[Fe/H]} \le +0.25$.
    \end{minipage}

  \end{center}
\end{table}

With the help of these model predictions we estimate that the slopes of
the observed \mgbp\/-- and \fep\/--$\sigma$ relations translate into a
change of $1.56 \pm 0.14$ and $0.40 \pm 0.12$ dex in metallicity per
dex $\log \sigma$ respectively (this assumes the models are error
free). In Figure~\ref{fig:mg_fe_sig1}b the long-dashed line represents
the slope of the \fep\/--$\sigma$ relation which would be expected for
a 1.56 dex change in metallicity, \ie\/ a slope of 0.083 in
equation~\ref{eq:metal_sig3}. Clearly the metallicity trend as
estimated from the \fep--$\sigma$ slope does not agree with the trend
seen in the \mgbp--$\sigma$ relation. This discrepancy shows of course
that one of our assumptions is wrong, \eg\/ is there evidence for a
significant age trend in index-$\sigma$ relations? A possible
explanation why the Fe--$\sigma$ relation is so shallow compared to the
Mg--$\sigma$ relation will be addressed in Sections~\ref{sec:age} and
\ref{sec:imp_sig}.

Besides the possibility of an age {\em trend}\/ along the sequence, it
has been suggested that the scatter at a given $\sigma$ should not be
interpreted so simply as in the above analysis. Trager et~al. (2000b)
concluded that it is possible to hide a large spread in stellar
populations at a given mass in a tight Mg--$\sigma$ relation, if ages
and metallicities are anti-correlated. For example, the weaker metal
lines due to a young stellar population can be balanced by an increase
in metallicity. In the literature one can indeed find several examples
of galaxies with young stellar populations showing a relatively high
metal content; notably, there are several such galaxies in the \gon\/
(1993) sample. Does our sample of galaxies also hide a significant
fraction of young galaxies in the Mg--$\sigma$ relation due to an
anti-correlation of age and metallicity effects? We explore this issue
further in Section~\ref{sec:age}.

\section{Ages, Metallicities and non-solar abundance ratios}
\label{sec:age}
In this section we investigate the age- and metallicity-distribution of
our sample with the help of age/metallicity diagnostic diagrams.
Particular emphasis is given to the effects and treatment of non-solar
abundance ratios. Using this new information we address in
Sections~\ref{sec:imp_index} and \ref{sec:imp_sig} two key questions:
(a) What combination of stellar population parameters (age,
metallicity, abundance ratios) leads to the observed slopes of the Fe--
and Mg--$\sigma$ relations? (b) Is there an age/metallicity
anti-correlation at work which hides a large spread of ages and
metallicities, at a given velocity dispersion, within tight
index--$\sigma$ relations?

\subsection{Initial age and metallicity estimates}
\label{sec:initial_age}
As emphasized by Worthey (1994), the determination of the ages and
metallicities of old stellar populations is complicated by the similar
effects that age and metallicity have on the integrated spectral energy
distributions. Broad-band colours and most of the line strength indices
are degenerate along the locus of $\Delta\, \rm{age} \simeq -3/2\,
\Delta\, \rm{Z}$. In the optical wavelength range only a few
narrow-band absorption line-strength indices have so far been
identified which can break this degeneracy. In terms of age, the Balmer
lines (H$\beta$, H$\gamma$ and H$\delta$) are the most promising
features, being clearly more sensitive to age than metallicity.
Absorption features like \mgb\/ \& \fe\/ are more metal sensitive. By
plotting an age-sensitive index and a metallicity-sensitive index
against each other one can (partly) break the degeneracy and estimate
luminosity-weighted age and metallicity of an integrated stellar
population \cite{gon93,fis96,meh98,jor99,kun2000,tra2000b}. The
usefulness of the \Hb\/ feature is limited by its sensitivity to
nebular emission and we would prefer to use a higher order,
emission-robust, Balmer line such as \Hg\/ \cite{kun2000}. However, due
to our restricted wavelength range we can only measure \Hb.

\begin{figure}
\epsfig{file=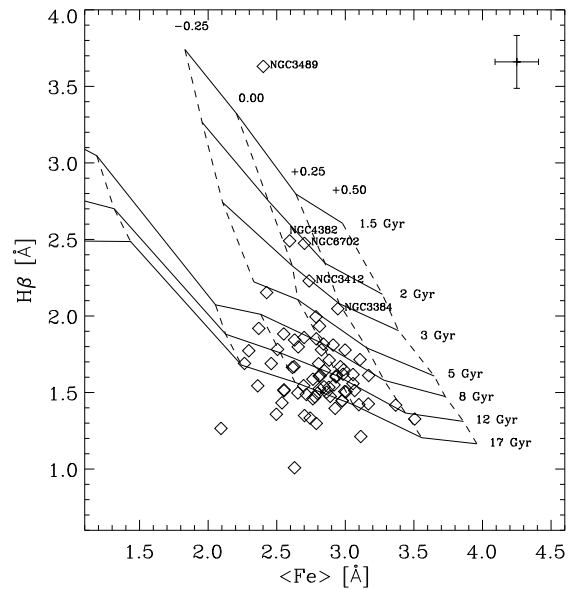,width=8.4cm}
\caption[]{\label{fig:fe_hb} Age/metallicity diagnostic diagram (\Hb\/
  {\em vs}\/ \fe\/ equivalent width) where \Hb\/ is corrected for
  nebular emission (see text for details). Models by Worthey (1994) are
  overplotted. The solid lines represent lines of constant age whereas
  the dashed lines are lines of constant metallicity. The steps in
  [Fe/H] are indicated at the top of the model grid, while the age
  steps are labeled at the right hand side. A median error bar for the
  sample is shown in the upper right corner.}
\end{figure}

Figure~\ref{fig:fe_hb} shows an age/metallicity diagnostic diagram
using \fe\/ as metallicity indicator and the emission corrected \Hb\/
as age indicator. Overplotted are models by Worthey (1994).\footnote{We
  refer here and in the following to Worthey~(1994), although we have
  used a later version of the models, as available May 2000 from Dr. G.
  Worthey's WWW page (single burst models, Salpeter IMF, Y=0.228 +
  2.7Z).} The majority of the galaxies occupy the region below the
8~Gyr line and show a spread in metallicity. A small number of galaxies
(NGC~3489, NGC~4382, NGC~6702, NGC~3412, and NGC~3384) with strong
\Hb\/ absorption indicating luminosity weighted ages $\le 5$~Gyrs is
also present, as well as a significant number of galaxies with
\Hb-absorption below the model prediction for 17~Gyrs. As we have
corrected the \Hb\/ index for emission, it is unlikely that most of
these galaxies are still affected by nebular emission \Hb\/ fill-in.
Some of the data points with low \Hb\/ absorption could be explained as
being scattered from the mean metallicity sequence due to Poisson noise
in our index measurements (see Section~\ref{sec:imp_index} for
corresponding Monte Carlo simulations). A further issue here are
non-solar abundance ratios which will be explored in the next section.

The overall distribution of the cluster/group dominated early-type
galaxies in this sample is reminiscent of the findings of Kuntschner \&
Davies (1998) and Kuntschner (2000) for the Fornax cluster. They
conclude that all elliptical galaxies follow a metallicity sequence at
roughly constant age. Particularly interesting is the paucity of
luminous, metal rich {\em and}\/ young galaxies in our sample, since
other samples of early-type galaxies such as \gon\/ (1993, see also
Trager et~al. 2000b, mainly field galaxies) and Longhetti (1998, shell
and pair galaxies) show a large relative fraction of these galaxies.

\subsection{Effects of non-solar abundance ratios}
\label{sec:effects}
How secure are the model predictions for the integrated light of
stellar populations? For a recent investigation of the uncertainties
see Trager et~al. (2000a). Here we would like to concentrate on one of
the most important systematic effects: {\em non-solar abundance
  ratios}\/ (hereafter NSAR). Over the last decade there has been a
growing consensus that the stellar populations of (luminous) elliptical
and lenticular galaxies show NSAR. Most notably magnesium, as measured
by the Mg$_2$ and \mgb\/ indices, compared to Fe, as measured in
various Fe-indices, does not track solar abundance ratio model
predictions, \ie\/ {${\rmn [Mg/Fe]} > 0$}
\cite{oco76,pel89,wor92,wei95,tan98,wor98,jor99,kun2000}.

Most of the currently available stellar population models cannot
predict the strength of indices as a function of [Mg/Fe], they are
limited to solar abundance ratios. This can lead to seriously flawed
age/metallicity estimates. For example, if in the presence of NSAR,
\mgb\/ is used as a metallicity indicator, the mean inferred ages are
younger and the mean metallicities are larger than what would be the
case if \fe\/ is used (see \eg\/ Worthey 1998, Kuntschner 2000). For
instance, if we take a 12 Gyr, solar metallicity stellar population and
correct it artificially to ${\rmn [Mg/Fe]} = +0.3$ (Trager et~al.
2000b, see also Table~\ref{tab:corr}), we get the following age and
metallicity estimates with respect to solar abundance ratio models: a
\fe\/ {\em vs}\/ \Hb\/ diagram gives an age of 13.4 Gyr and ${\rmn
  [Fe/H]} = -0.11$ whereas in a \mgb\/ {\em vs}\/ \Hb\/ diagram we
estimate an age of 7.7 Gyr and ${\rmn [Fe/H]} = +0.22$. This example
shows that if NSAR are not properly treated they lead to wrong
age/metallicity estimates which are correlated such that an
overestimated metallicity leads to younger ages and vice versa. We
note, that if a diagram of [MgFe]\footnote{${\rmn [MgFe]} = \sqrt{{\rmn
      Mg}\,b\, \times <{\rmn Fe}>}$} {\em vs}\/ \Hb\/ is used then
abundance ratio effects are reduced and an age of 11.3 Gyr and
[Fe/H]=0.02 are estimated. As the NSAR will seriously affect our age
and metallicity estimates we will describe in the next few paragraphs
how we correct for them and re-analyse the corrected data in
Sections~\ref{sec:imp_index} and \ref{sec:imp_sig}.

\begin{figure}
  \epsfig{file=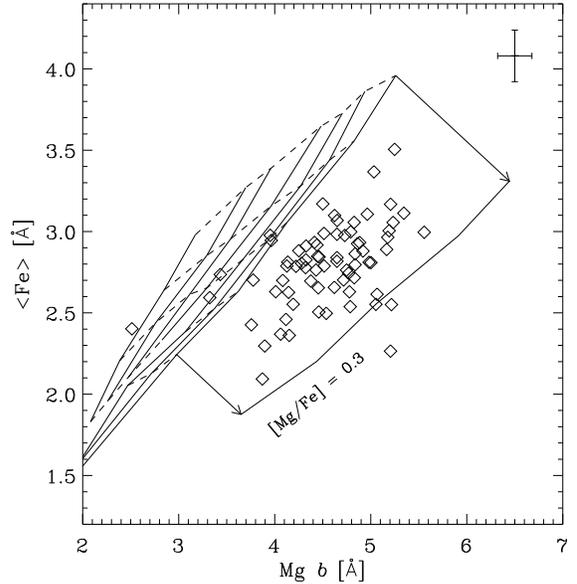,width=8.4cm}
\caption[]{\label{fig:mgb_fe} \mgb\/ {\em vs}\/ \fe\/
  diagram. Overplotted are models by Worthey (1994).  The solid lines
  represent lines of constant age whereas the dashed lines are lines of
  constant metallicity. A correction for non-solar abundance ratios of
  Mg and Fe is shown for the 17 Gyr line (arrows; Trager~et~al. 2000a,
  model~4). The median observational errors are indicated in the right
  upper corner.}
\end{figure}

The influence of NSAR can be examined in a Mg-index {\em vs}\/ Fe-index
plot \cite{wor92}. In such a diagram the model predictions (at solar
abundance ratios) cover only a narrow band in the parameter space as
effects of age and metallicity are almost degenerate.
Figure~\ref{fig:mgb_fe} shows our sample in such a diagram.
Overplotted are model predictions from W94. The great majority of the
galaxies do not agree with the model predictions, a result which is
generally interpreted as the effect of NSAR. Previous authors
\cite{wor92,jor99,kun2000} found a trend in the sense that the more
luminous galaxies show greater ratios of [Mg/Fe]. In
Figure~\ref{fig:mgb_fe} this trend is not very pronounced but see
Section~\ref{sec:imp_sig} where we discuss the trend of [Mg/Fe] {\em
  vs}\/ $\sigma$.

Recently Trager~et~al. (2000a) investigated the effects of NSAR in
early-type galaxies for the sample of galaxies from \gon's thesis
(1993). They tabulated corrections for various scenarios of NSAR for a
selection of important indices in the Lick/IDS system based on the
calculations by Tripicco~\&~Bell (1995) and Worthey models. We use
their best fitting model (model~4, see also Table~\ref{tab:corr}) to
correct our index measurements to solar abundance ratios and then
derive {\em improved}\/ age/metallicity estimates with respect to W94
models. Trager et~al. corrected the stellar population models to fit
each galaxy individually; here we have chosen instead to correct the
index measurements rather then the models in order to present
age/metallicity diagnostic diagrams with all galaxies plotted on a
common model.

The corrections given by Trager~et~al. are certainly an important step
forward to derive better age/metallicity estimates but are not to be
taken as final because \eg\/ the corrections are given under the
assumption that at fixed {\em total}\/ metallicity one can scale the
solar-ratio isochrones and compute integrated stellar population
models. This assumption has been challenged by Salaris \& Weiss (1998)
who predict hotter isochrones for $\rmn{[Mg/Fe]} > 0.0$ (calculated at
sub-solar metallicities).

The corrections given by Trager~et~al. (see Table~\ref{tab:corr})
indicate that for [Mg/Fe]$> 0.0$ the \mgb\/ index increases while {\em
  at the same time}\/ the \fe\/ index decreases compared to solar
abundance ratios. A correction for a 17~Gyr model prediction to
$\rmn{[Mg/Fe]} = +0.3$ (typical for the most luminous galaxies) is
shown in Figure~\ref{fig:mgb_fe}. A qualitatively similar correction
for the index combination of Fe5270 and Mg$_2$ was previously published
by Greggio (1997).

\begin{table}
  \caption[]{Fractional index responses to non-solar abundance ratios}
  \label{tab:corr}
  \begin{center}

  \begin{tabular}{cllll} \hline
     Model & $\delta$\Hb  &$\delta$MG$_2$ &$\delta$\mgb&$\delta$\fe \\ 
       4   & 0.027        &0.086          &0.225       &$-0.164$      \\ \hline         
    \end{tabular}

  \medskip
  \begin{minipage}{8.0cm}
    Notes: Fractional index responses for $\Delta$[Fe/H]=-0.3 dex at
    [Z/H]=0, in the sense $\delta I = \Delta I/I$, where $\Delta I$ is
    the index change and $I$ is the original values of the index. Both
    C and O are enhanced. The values are taken from
    Trager~et~al.~(2000a).
  \end{minipage}

\end{center} 
\end{table}

Using the distribution of the galaxies in the \mgb\/ {\em vs} \fe\/
diagram and the (inverse) corrections given in Table~\ref{tab:corr} we
calculated for each galaxy the \fe\/ strength and \mgb\/ strength it
would have at {\em solar abundance ratios}. Graphically this can be
seen as shifting each galaxies individually along a vector with the
same direction until it meets the solar abundance ratio models. In a
\mgb\/ {\em vs}\/ \fe\/ diagram the models are not completely
degenerate in age and metallicity so we have to take the age of the
galaxies into account.  For this purpose we used the ages derived from
a \Hb\/ {\em vs}\/ \fe\/ diagram.  The age estimates were improved
through an iterative scheme, in which the second and third iterations
employ the \Hb\ and \fe\/ values corrected for NSAR. This procedure was
tested on the \gon\/ (1993) sample giving the same results as the
procedure used by Trager~et~al. (2000a) within an accuracy of $\delta
\rmn{[Mg/Fe]} = \pm 0.02$. The indices \Hb\/ and Mg$_2$ were also
corrected using the overabundance estimates from the \mgb\/ {\em vs}\/
\fe\/ diagram. The mean change in the \Hb\/ index is very small with
$\Delta index = -0.03$~\AA, whereas the mean Mg$_2$ index has changed
by --0.015~mag.  We note, that deriving the [Mg/Fe] ratios from a
Mg$_2$ {\em vs}\/ \fe\/ diagram leads to very similar results.

\begin{figure}
\epsfig{file=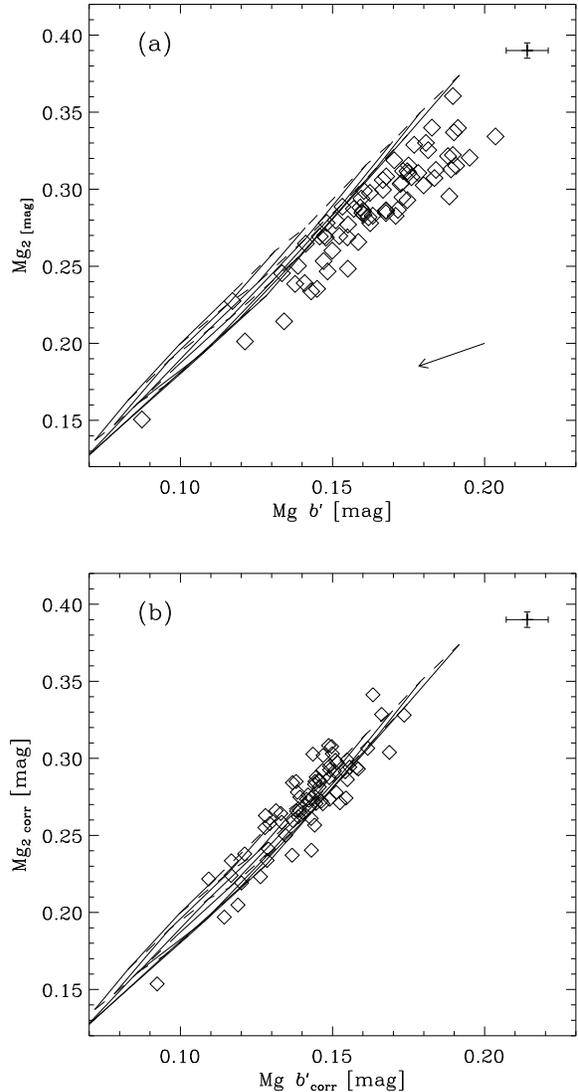,width=8.4cm}
\caption[]{\label{fig:mg2_mgb} (a) Mg$_2$ {\em vs}\/ \mgbp
  relation. (b) Mg$_2$ {\em vs}\/ \mgbp relation corrected for
  non-solar abundance ratios. The arrow in panel (a) indicates the mean
  correction for non-solar abundance ratios.}
\end{figure}

As Trager~et~al. emphasize, it is not really the enhancement of
magnesium which increases the \mgb\/ or Mg$_2$ index for
`Mg-overabundant galaxies' but rather a deficit of Fe (and Cr).  Hence
they propose to change the terminology from `Mg-overabundance' to
`Fe-peak element deficit'. In fact, all Lick/IDS indices are affected
not only by one element, such as magnesium or Fe, but by a complex
combination of all the species contributing to the three bandpasses.
For example the indices Mg$_2$ and \mgbp\/ are both dominated by
magnesium but have slightly different sensitivities towards Mg and
other elements \cite{tri95} as can be seen in the following example.
Wegner et~al. (1999) found for the EFAR sample a tight correlation
between the \mgbp\/ index and Mg$_2$, but when compared to the model
predictions this relation is in significant disagreement. In
Figure~\ref{fig:mg2_mgb}a we show the Mg$_2$ {\em vs}\/ \mgbp\/
relation for our sample. When we correct both indices for non solar
abundance ratios they are brought into excellent agreement with the
models (Figure~\ref{fig:mg2_mgb}b).

\section{Improved age and metallicity estimates}
\label{sec:imp_index}
\begin{figure*}
\epsfig{file=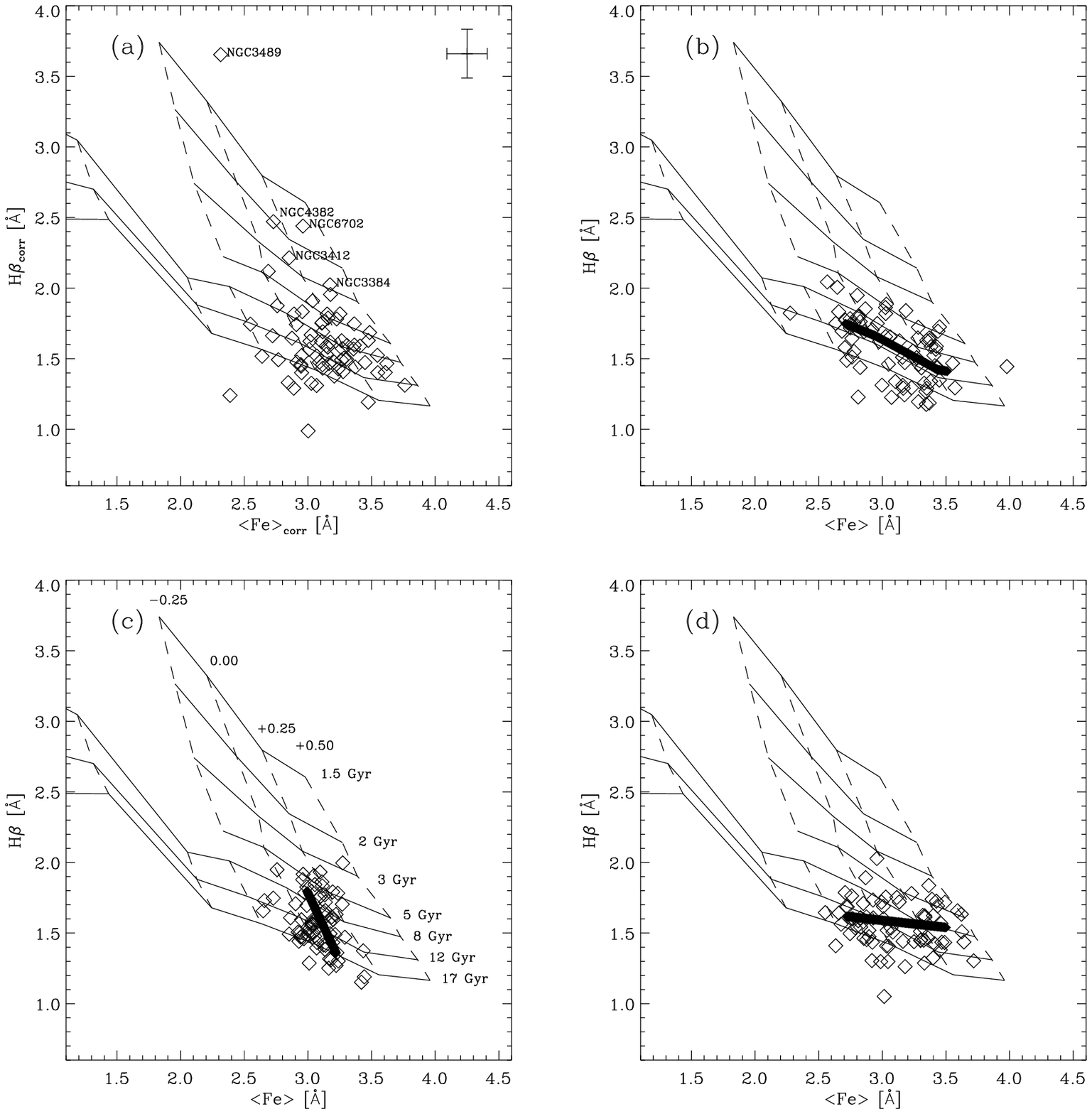,width=17.7cm}
\caption[]{\label{fig:fe_hb2} (a) \Hb\/ {\em vs}\/
  \fe\/ equivalent width diagram corrected for non-solar abundance
  ratios (see text for details). Models by Worthey (1994) are
  overplotted (see panel (c) for labels). The solid lines represent
  lines of constant age whereas the dashed lines are lines of constant
  metallicity. Panel (b) shows a Monte Carlo simulation (open diamonds)
  of a galaxy sample spanning a range in metallicity ($-0.12 \le {\rmn
    [M/H]} \le +0.31$) at constant age (10.7 Gyr, filled diamonds). The
  errors used in the simulation reflect our median observational
  errors. As alternative hypotheses we also constructed a galaxy sample
  with a sequence in age ($7.1 \le age \le 15.1$~Gyr) at constant
  metallicity (${\rmn [M/H]} = +0.11$) and a galaxy sample with
  correlated variations in both age ($7.8 \le age \le 14.6$) and
  metallicity ($-0.17 \le {\rmn [M/H]} \le 0.37$), which are shown in
  panels (c) \& (d) respectively.}
\end{figure*}

Having corrected our indices for NSAR we can re-examine the
age/metallicity distribution. The inferred metallicities are now a
better estimation of the {\em total}\/ metallicity, hereafter [M/H],
rather than being biased towards a specific element.
Figure~\ref{fig:fe_hb2}a shows an age/metallicity diagnostic diagram
using the NSAR-corrected values of \Hb\/ and \fe, denoted by the
underscript `corr'. The mean \fe\/ absorption increased from 2.79($\pm
0.25$)~\AA\/ to 3.11($\pm 0.28$)~\AA\/ and the mean \Hb\/ absorption
changed from 1.61($\pm 0.22$)~\AA\/ to 1.58($\pm 0.21$)~\AA\/ after the
correction. In order to avoid bias by outliers the former means were
calculated using a 3$\sigma$ clipping algorithm. The difference in
\Hb\/ absorption strength is negligible but clearly the change in \fe\/
causes a significant change in the metallicity {\em and}\/ age
estimates leading to overall younger ages and higher metallicities. At
the same time the metallicity sequence seems to be preserved if not
more pronounced. The number of galaxies with weak \Hb-absorption
suggesting ages in excess of 17~Gyr is slightly reduced, but not
completely eliminated.

We already speculated in Section~\ref{sec:initial_age} that these `old'
galaxies maybe scattered from the metallicity sequence due to Poisson
noise in the index-measurements. In order to test this hypothesis, we
have performed simple Monte Carlo (hereafter MC) simulations. In
Figure~\ref{fig:fe_hb2}b we show a mock sample with a metallicity
sequence ($-0.12 \le {\rmn [M/H]} \le +0.31$) at constant age (10.7
Gyr, filled diamonds). The galaxies are uniformly distributed along the
sequence. Each `galaxy' in this sample is then perturbed with our
median observational errors and the resulting distribution of a
representative simulation is shown as open diamonds. The input sequence
to the MC simulation was constructed such that the mean and 1$\sigma$
scatter of the \Hb\/ and \fe\/ distribution of the real data is
matched. The chosen sequence of galaxies at constant age closely
resembles our observed distribution once the observational errors are
taken into account. Most of the galaxies with low \Hb\/ absorption are
also consistent with being scattered due to errors. However, the few
galaxies with very strong \Hb\/ absorption in our observed sample are
not reproduced in the MC simulations and are therefore likely to have
genuinely younger stellar populations. We note, that most of these
young galaxies are not in clusters as three of them are in the Leo
group, one resides in a low density environment and one only is in the
Virgo cluster.

As alternative hypotheses we also constructed a galaxy sample with a
sequence in age ($7.1 \le age \le 15.1$~Gyr) at constant metallicity
(${\rmn [M/H]} = +0.11$) and a galaxy sample with anti-correlated
variations in age ($7.8 \le age \le 14.6$) and metallicity ($-0.17 \le
{\rmn [M/H]} \le 0.37$), which are shown in Figure~\ref{fig:fe_hb2}c
and d respectively.  Again we tried to match the mean and scatter of
the observed sample as well as possible.

The simulation with constant input metallicity reproduces well the
distribution in \Hb\/ but we cannot reproduce the spread in \fe\/ at
the same time (mock sample spread: 0.17~\AA\/ as opposed to 0.28~\AA\/
for the observed sample). Therefore we conclude that our observed
sample shows genuine spread in metallicity and reject the hypothesis of
constant metallicity. We note, that under the extreme assumption of
constant metallicity the mean age of the mock sample is as old as
11.0~Gyrs supporting our earlier claim that most of the galaxies are
old.

Whether our sample follows an age-metallicity anti-correlation or not
is more difficult to address. Overall, the simulation reproduces the
observed distribution quite well. This is not surprising as the input
ages for the MC simulation range from $\sim$8 to 14~Gyrs where the age
discrimination power of the \Hb\/ index is very limited.
  
Using a stronger age-metallicity anti-correlation as input sequences
for the MC simulations leads readily to disagreements between the
overall distribution of the observed sample and the simulations.
Pinpointing the exact slope where the input model is in disagreement
with the data depends sensitively on the statistical method applied to
the data. A more thorough treatment of this issue requires also a more
sophisticated model hypothesis, \eg\/ a non-uniform distribution of
galaxies along the sequence and is beyond the scope of this paper.

In summary we conclude that our sample of 72 early-type galaxies
contains only a small number ($\sim$5) of galaxies with mean luminosity
weighted ages of $\le 3$~Gyr. The main body of the data is, within our
measurement errors, consistent with either a constant age sequence at
$\sim$11~Gyr or a mild age-metallicity anti-correlation with ages $\ge
8$~Gyr (such as the one shown in Figure~\ref{fig:fe_hb2}d).

\begin{figure*}
\epsfig{file=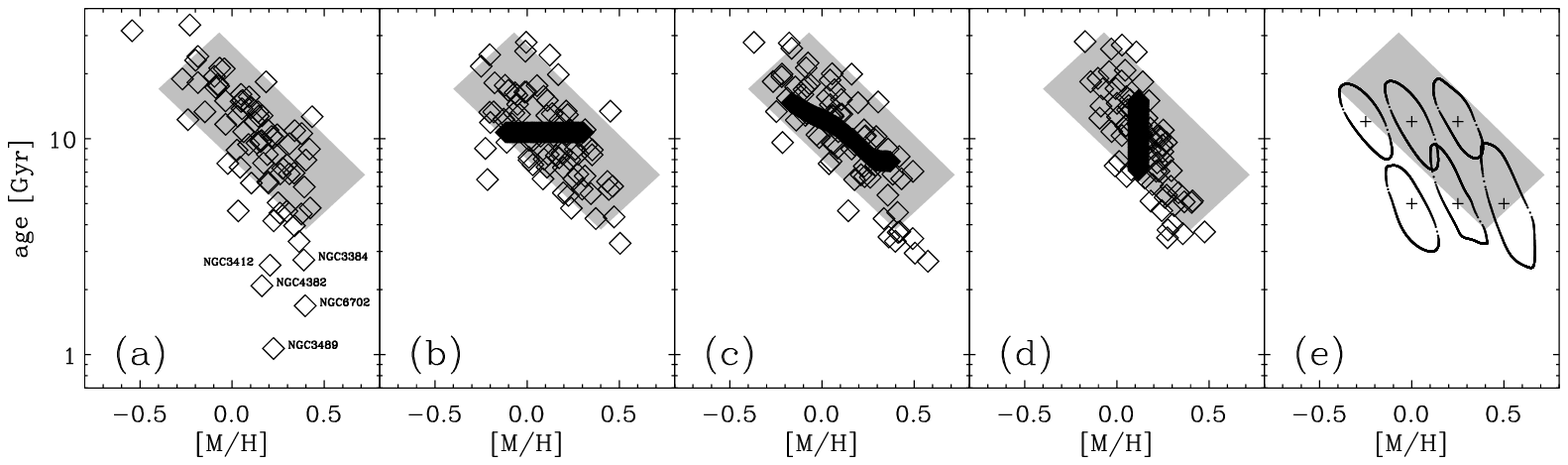, width=17.5cm}
\caption[]{\label{fig:age_metal_sim} Age {\em vs}\/ metallicity 
  distributions: (a) our sample of 72 galaxies as derived from
  Figure~\ref{fig:fe_hb2}a, (b) a Monte Carlo simulation with constant
  age and varying metallicity. The filled diamonds represent the input
  data set for the simulations, (c) a Monte Carlo simulation with
  varying age and varying metallicity, (d) a Monte Carlo simulations
  with constant metallicity, (e) error contours for selected points in
  the age-metallicity plane based on our median observational errors.
  The shaded area is the same in all diagrams and can be used to
  compare the distributions.}
\end{figure*}

We can learn more from the MC simulations when we actually measure the
ages and metallicities from the diagrams in Figure~\ref{fig:fe_hb2}.
This was done by interpolating between the model grid points and also
linearly extrapolating if data points are outside the model
predictions.

In the age/metallicity plane (Figure~\ref{fig:age_metal_sim}a) our
observed sample follows a trend, such that more metal rich galaxies are
also younger. This trend has recently been discussed by several authors
(\eg\/ J{\o}rgensen 1999, Trager et~al. 2000b, Terlevich \& Forbes
2000).  However, analysing the MC simulation in
Figure~\ref{fig:age_metal_sim}b we find that even a metallicity
sequence at {\em constant age}\/ is transformed into an age-metallicity
correlation due to correlated errors in age/metallicity diagnostic
diagrams such as Figure~\ref{fig:fe_hb2}. For comparison we show in
Figure~\ref{fig:age_metal_sim}c a MC simulation with both, age and
metallicity variations and in Figure~\ref{fig:age_metal_sim}d a MC
simulation with constant metallicity as input.  The correlated errors
stem from the fact that an error in \eg\/ the metallicity index results
an error in both, the age {\em and}\/ metallicity estimate, such that
if the age is underestimated the galaxy seems also more metal rich and
vice versa. Error contours in the age/metallicity plane, based on our
median observational errors, for selected ages and metallicities are
shown in Figure~\ref{fig:age_metal_sim}e. The contours are elongated
mainly in the direction of a negative correlation of age and
metallicity, however the detailed shape depends on the exact position
within the age/metallicity plane.  We note, that the error contours in
Figure~\ref{fig:age_metal_sim}e reflect only the observational errors
in our index measurements and do not include any errors in our
determination of [Mg/Fe]. As demonstrated in section~\ref{sec:effects}
they do also lead to an artificial age/metallicity correlation.

The effects of correlated errors will be present in diagrams such as
Figure~\ref{fig:fe_hb2} as long as the index-combination does not
completely resolve the age/metallicity degeneracy. Therefore any
findings based on these diagrams such as age-metallicity correlations
depend crucially on the size of the errors in the observables. Ideally
one would prefer errors in \Hb\/ $< 0.1$~\AA, however, large samples
such as the Coma compilation by J{\o}rgensen (1999) and the Lick
extragalactic database \cite{tra98} show typical errors of 0.2 to
0.3~\AA\/ with a tail of even larger errors making a proper
age/metallicity analysis impossible. Our sample has a median error of
0.17~\AA\/ in \Hb\/ with the largest being 0.23~\AA\/ and therefore is
perhaps just at the borderline where one can start a useful
age/metallicity analysis.

We note, that our sample contains a small number of galaxies with young
stellar populations which tend to be more metal rich than the average
galaxy. Looking back at the Mg--$\sigma$ relation shown in
Figure~\ref{fig:mg_fe_sig1}a only NGC~3489 and NGC~4382 do clearly
deviate towards lower \mgbp\/ values, the other galaxies with young
stellar populations are consistent with the main relation which in turn
can be perhaps best explained by a negative correlation of age and
metallicity which holds these galaxies on the main relation (see also
Figure~\ref{fig:age_metal_sig2}).

For the remaining galaxies in our sample we do not find clear evidence
of an age-metallicity anti-correlation and therefore conclude that the
scatter at a given $\sigma$ in the \mgbp--$\sigma$ relation is not
significantly reduced due to an age-metallicity `conspiracy'.

\section{Corrected index--$\sigma$ relations}
\label{sec:imp_sig}
\begin{figure}
\epsfig{file=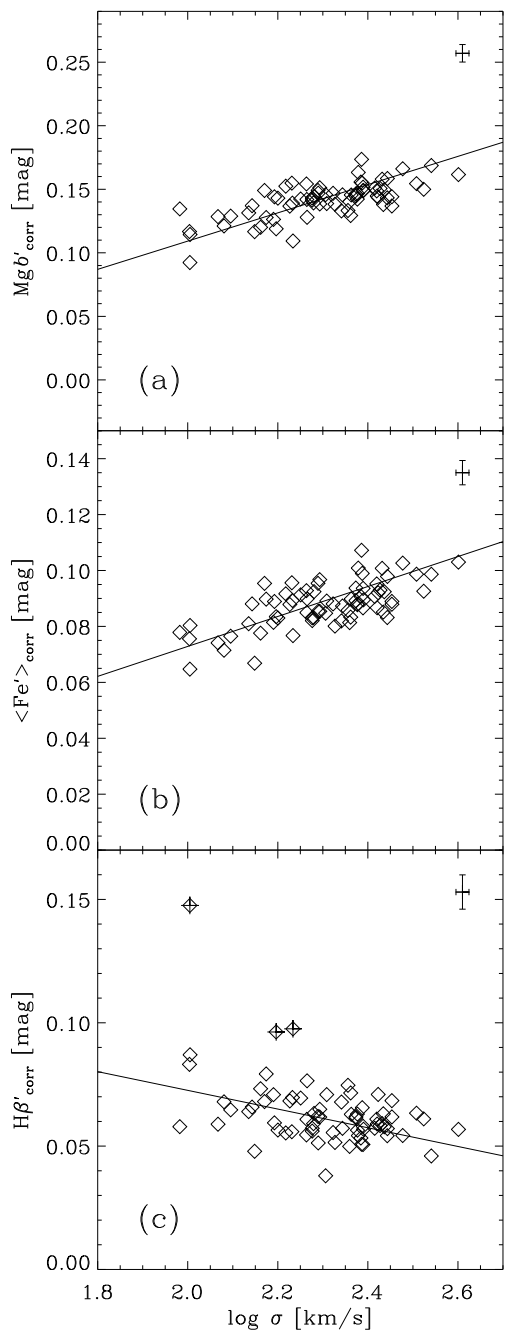,width=8.4cm}
\caption[]{\label{fig:mg_fe_sig2} (a) The \mgbp\/--$\sigma$, (b)
  \fep\/--$\sigma$ and (c) \hbp\/--$\sigma$ relations corrected for
  non-solar abundance ratios. The solid lines indicate a straight line
  fit taking into account error in both variables. The plus signs in
  panel (c) indicate galaxies which were excluded from the fit via an
  iterative scheme, see text for details. Error bars representing the
  median errors in each quantity are shown in the upper right corner of
  each panel.}
\end{figure}

Of course the correction of the \mgb\/ and \fe\/ indices for NSAR will
also affect the index--$\sigma$ relations. The corrected versions are
shown in Figure~\ref{fig:mg_fe_sig2}. Here we find `corrected' fits
({\tt fitexy} routine) of

\begin{equation}
  \label{eq:mgb}
  {\rmn Mg{\,\it b}}\,^\prime _{\rmn corr}= 0.091(\pm0.010) \log \sigma - 0.068(\pm0.023) \, ,
\end{equation}

and

\begin{equation}
  \label{eq:fe}
 <{\rmn Fe\,^\prime}> _{\rmn corr} = 0.044(\pm0.006) \log \sigma - 0.012(\pm0.014) \, .
\end{equation}

The scatter about the \mgbp$_{\rmn corr}$--$\sigma$ relation is reduced
to 0.010. However, the corrections of the \mgb\/ index for NSAR are
based on the \mgb\/ {\em vs}\/ \fe\/ diagram thus correlated errors are
again present and we cannot draw any firm conclusions about the reduced
scatter when we take the size of our observational errors into account.

Compared with the un-corrected versions, the \mgbp$_{\rmn
  corr}$--$\sigma$ relation shows a shallower slope whereas the
\fep$_{\rmn corr}$--$\sigma$ relation is steeper. This may seem to be
an obvious result of the NSAR corrections, nevertheless it is important
as the following shows. Using the W94 models and assuming that there
are no age variations along the metal index--$\sigma$ relations, one
can translate the slopes into changes of [M/H], giving $\Delta {\rmn
  [M/H]}/ \Delta \log \sigma = 1.00 \pm 0.11$~dex and $= 0.83 \pm
0.11$~dex for \mgbp\/ and \fep\/ respectively (see also
Table~\ref{tab:indexchange}). This shows that the corrected
index--$\sigma$ relations give now consistent measurements of the
(total) metallicity change with central velocity dispersion, \ie\/
there is no need to invoke an additional age trend along the sequence
after the correction for NSAR. We conclude that the disagreement in the
strength of the metallicity change with $\log \sigma$, as estimated
from the \fep-- and \mgbp--$\sigma$ relations, can be resolved if NSAR
are taken into account.

As a consistency test, we show in Figure~\ref{fig:mg_fe_sig2}c the
\hbp$_{\rmn corr}$--$\sigma$ relation. If there is no age trend along
the index--$\sigma$ relations, then the slope of our most age-sensitive
index should only reflect the change in metallicity along the sequence.
Towards the low $\sigma$ range we can see the galaxies with strong
\Hb\/ absorption deviating from the main \Hb--$\sigma$ relation.  Due
to these strong deviations it is difficult to establish a fit which
represents the overall slope. We therefore used an iterative scheme
where we excluded all galaxies deviating by more than 3$\sigma$ from
the fit (2 iterations, indicated in Figure~\ref{fig:mg_fe_sig2}c by
plus signs). The best fitting relation is:

\begin{equation}
  \label{eq:hb1}
  {\rmn H} \beta \,^\prime _{\rmn corr} = -0.031(\pm 0.008) \log \sigma + 0.133(\pm 0.019) \, .
\end{equation} 

Using the W94 stellar population models and assuming no age trend along
the sequence the fit translates into $\Delta {\rmn [M/H]}/ \Delta \log
\sigma = 1.00 \pm 0.26$~dex, which is in good agreement with the above
results.

If we assume an age-metallicity anti-correlation such as the one shown
in Figure~\ref{fig:fe_hb2}d ($7.8 \le age \le 14.6$, $-0.17 \le {\rmn
  [M/H]} \le 0.37$) the models predict a slope in the \Hb--$\sigma$
relation of --0.005, which is close to no change in \Hb\/ with central
velocity dispersion. This is inconsistent with our data on the $3-4
\sigma$ level. Taking the 1$\sigma$ error of the observed \Hb--$\sigma$
slope into account, allows only for an age gradient of 2 to 3 Gyrs
along the sequence.

We conclude that the slope of the \hbp$_{\rmn corr}$--$\sigma$ relation
is consistent with the other index--$\sigma$ relations and our
assumption of no age trend along the index--$\sigma$ relations within
the fitting errors. Although there are some individual galaxies which
show signs of an age-metallicity anti-correlation which can reduce the
spread of scaling relations at a given mass, we do not favour a picture
where such a correlation acts {\em along} the scaling relations.

We emphasize that although the total metallicity increases by $0.9 \pm
0.1$~dex per dex in $\sigma$, individual elements can contribute
differently to the increase. For example Fe seems to change only very
little whereas magnesium (and probably other $\alpha$-elements) show a
steeper increase.

\begin{figure}
\epsfig{file=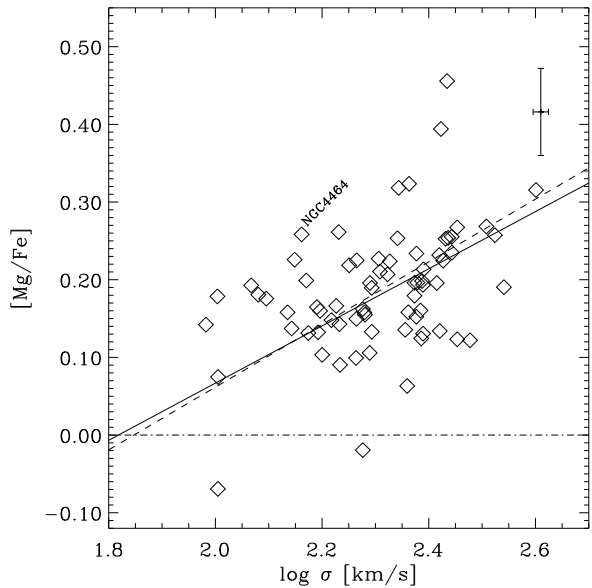,width=8.4cm}
\caption[]{\label{fig:mgfe_sig} [Mg/Fe] {\em vs}\/ $\log \sigma$
  relation. The median observational errors are indicated in the right
  upper corner. The solid line shows a straight line fit taking into
  account errors in both variables. The dashed line represents the best
  fitting relation to the sample of Trager et~al. (2000b).}
\end{figure}

In agreement with previous investigations into the [Mg/Fe] ratios
\cite{wor92,jor99,kun2000,tra2000b} we find that the [Mg/Fe] ratio
slowly increases with the central velocity dispersion (or luminosity).
Using the method of Trager et~al. to measure the NSAR we estimate $0.05
\le \rmn{[Mg/Fe]} \le 0.30$ in the range $100 \le \sigma \le 400$
(Figure~\ref{fig:mgfe_sig}). The largest [Mg/Fe] ratios are
approximately 0.1 dex lower than what was estimated with previous
methods \cite{wor92,wei95,jor99,kun2000}. We note, that towards the low
velocity dispersion range of our data, early-type galaxies still show
{\em non-solar} abundance ratios suggesting that early-type galaxies
with solar abundance ratios, if they exist at all, have velocity
dispersions below $\sim$100~\kms.

We further note, that there are low luminosity galaxies exhibiting
[Mg/Fe] ratios similar to much more luminous galaxies. The best
confirmed case here is NGC~4464 \cite{pel99,vaz2000}. This in turn
suggests that the [Mg/Fe]--$\sigma$ relation may show a significant
scatter which would be responsible for some of the scatter in the
Mg--$\sigma$ relation. The best fitting [Mg/Fe]--$\sigma$ relation is

\begin{equation}
  \label{eq:mgfe_sig}
  {\rmn [Mg/Fe]} = 0.30(\pm0.06) \log \sigma - 0.52(\pm0.15) \, .
\end{equation}

A (non-parametric) Spearman rank-order test shows a correlation
coefficient of 0.40 with a significance level of $<0.1$\%. The rather
small correlation coefficient indicates that this relation shows a lot
of scatter. However, it is again very sensitive to selection effects at
the low $\sigma$ end which is not well represented in our data set.
The relation found for our sample is in excellent agreement with the
results of Trager et~al. (2000b; Fig.~\ref{fig:mgfe_sig}, dashed line).
We note, that J{\o}rgensen's (1999) analysis of Coma galaxies indicates
a much steeper slope of $\Delta {\rmn [Mg/Fe]}/ \Delta \log \sigma
\simeq 1.1$.

\begin{figure}
\epsfig{file=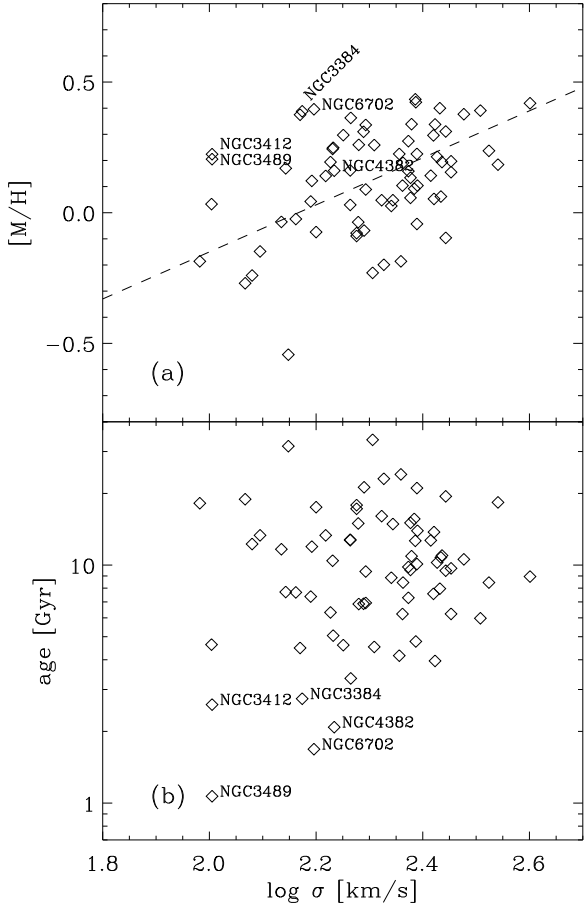,width=8.4cm}
\caption[]{\label{fig:age_metal_sig2}(a) The [M/H]--$\sigma$ relation
  and, (b) the $\log$~age--$\sigma$ relation for our sample of 72
  galaxies. The ages and metallicities are derived from
  Figure~\ref{fig:fe_hb2}a. The dashed line in panel (a) is not a fit
  to the data but represents a slope of 0.9 dex in metallicity per dex
  in $\log \sigma$.}
\end{figure}

Finally we want to present the correlations between our derived ages
and metallicities with the central velocity dispersion. We note, that
the derived parameters carry all the caveats mentioned in the previous
sections. Figure~\ref{fig:age_metal_sig2}a and b show the [M/H]-- \&
$\log$~age--$\sigma$ relations respectively. Although there is
substantial spread in the relations it can be clearly seen that our
derived metallicities are consistent with an increase of $\Delta {\rmn
  [M/H]}/ \Delta \log \sigma \simeq 0.9$ (indicated by the dashed line
in Figure~\ref{fig:age_metal_sig2}a). A (non parametric) Spearman
rank-order test shows a weak correlation coefficient of 0.33 with a
significance level of $0.4$\%. The change in metallicity per dex in
$\log \sigma$ is in good agreement with the relation found by
Kuntschner (2000) for the Fornax cluster and by Trager et~al. (2000b)
for early-type galaxies in groups and clusters. Our best age estimates
do not show a significant correlation with $\sigma$ (Spearman
rank-order test: correlation coefficient $= 0.10$, significance~level
$= 40$\%). However, we emphasize that the age spread increases towards
the low $\sigma$ end of the relation. For a similar result see
Concannon, Rose \& Caldwell (2000).

\section{CONCLUSIONS}
\label{sec:conclusions}

For our sample of 72 early-type galaxies, drawn mostly from clusters
and groups, we have analysed the \mgbp\/--$\sigma$ and \fep\/--$\sigma$
relations as well as age/metallicity diagnostic diagrams with
up-to-date stellar population models. Taking the effects of non-solar
abundance ratios into account, we conclude:

\begin{itemize}
  
\item Tight index-$\sigma$ relations, as well as the results from
  age/metallicity diagnostic diagrams, provide evidence that cluster
  early-type galaxies are made of old stellar populations (age
  $>$7~Gyrs) and follow mainly a trend of increasing metallicity with
  increasing central velocity dispersion $\sigma$. The galaxies display
  no or at most a mild age-trend along the sequence. A small number
  ($\sim$5) of galaxies show younger luminosity weighted ages and above
  average metallicities in their mean stellar populations.  These
  galaxies have velocity dispersions below $\sim$170~\kms.
  
\item Correcting the line-strength indices for non-solar abundance
  ratios is a crucial step forward in deriving more reliable
  age-metallicity estimates. Correlated errors, however, in
  age/metallicity diagnostic diagrams combined with the precision of
  currently available index measurements for medium to large samples
  lead to artificial anti-correlations between age and metallicity
  estimates. Therefore it is difficult to differentiate between mild
  anti-correlations between age and metallicity and a pure metallicity
  sequence. For the majority of our sample we do not find evidence for
  a prominent age/metallicity correlation and we can firmly exclude a
  pure age sequence at constant metallicity.
  
\item Correcting the \mgb\/ and \fe\/ indices for non-solar abundance
  ratios changes the slope of the index--$\sigma$ relations in the
  sense that the Mg--$\sigma$ relation becomes shallower and the
  \fe\/--$\sigma$ relation becomes steeper. The \Hb--$\sigma$ relation
  remains virtually unchanged. Under the assumption of no age gradients
  along the index--$\sigma$ relations all three index--$\sigma$
  relations give consistent estimates of the total metallicity change
  per dex in $\log \sigma$: $\Delta {\rmn [M/H]}/ \Delta \log \sigma =
  1.00\pm0.11$, $0.83\pm0.11$ and $1.00\pm 0.26$ for the
  \mgb--$\sigma$, $<$Fe$>$--$\sigma$, and \Hb--$\sigma$ relations
  respectively. Introducing an age gradient along the sequence leads
  readily to inconsistent results of model predictions and the slope of
  the \Hb--$\sigma$ relation. At most an age gradient of 2 to 3 Gyr
  along the scaling relations is within our measurement errors.

\item The [Mg/Fe] ratio is mildly correlated with the central velocity
  dispersion and ranges from ${\rmn [Mg/Fe]} = 0.05$ to 0.3 for
  galaxies with $\sigma \ge 100$~\kms. Some low luminosity galaxies,
  such as NGC~4464, do not follow the main trend but show [Mg/Fe]
  ratios similar to the more luminous galaxies.
     
\end{itemize}

\section*{ACKNOWLEDGEMENTS}
We thank Stephen Moore for providing us with a code to calculate error
contours. H.K. and R.J.S. were supported at the University of Durham by
a PPARC rolling grant in Extragalactic Astronomy and Cosmology. R.J.S.
also thanks Fondecyt-Chile for support through Proyecto FONDECYT
3990025. R.L.D. acknowledges a Leverhulme Trust fellowship and a
University of Durham Derman Christopherson fellowship.

\appendix
\section{Line strength indices}
Table~\ref{tab:table} lists the name, environment, central velocity
dispersion and line-strength indices for all 72 galaxies in our sample.

\begin{table*}
  \caption[]{Index Measurements}
  \label{tab:table}
  \begin{tabular}{llcccccccccccc} \hline 
  Name & env. & S/N & $\log \sigma_0$ & $\epsilon_{\sigma_0}$ & \Hb & $\epsilon_{\rmn{H}\beta}$ & Mg$_2$ &   $\epsilon_{\rmn{Mg}_2}$ & \mgb & $\epsilon_{\rmn{Mg}\/b}$ & \fe & $\epsilon_{<\rmn{Fe}>}$ & \OIIIb\\ \hline
   NGC~2300 &   A569 & 45 &  2.427 &  0.014 &   1.53 &   0.15 &  0.311 &  0.005 &   4.92 &   0.16 &   2.88 &   0.14 &    -  \\
   NGC~2320 &   A569 & 36 &  2.508 &  0.022 &   1.67 &   0.21 &  0.295 &  0.006 &   5.18 &   0.20 &   2.96 &   0.18 &  -1.98\\
   NGC~2329 &   A569 & 32 &  2.359 &  0.024 &   1.30 &   0.23 &  0.270 &  0.006 &   4.13 &   0.23 &   2.79 &   0.20 &  -0.37\\
   NGC~2340 &   A569 & 34 &  2.386 &  0.016 &   1.33 &   0.21 &  0.340 &  0.006 &   5.25 &   0.21 &   3.51 &   0.18 &    -  \\
   NGC~2513 &      - & 47 &  2.420 &  0.014 &   1.60 &   0.15 &  0.329 &  0.004 &   4.88 &   0.16 &   2.93 &   0.14 &    -  \\
   NGC~2634 &  Group & 40 &  2.276 &  0.014 &   1.50 &   0.18 &  0.286 &  0.005 &   4.45 &   0.18 &   2.65 &   0.16 &  -0.37\\
   NGC~2672 &  Group & 34 &  2.423 &  0.023 &   1.88 &   0.19 &  0.337 &  0.006 &   5.21 &   0.21 &   2.55 &   0.19 &    -  \\
   NGC~2693 &  Group & 32 &  2.541 &  0.022 &   1.21 &   0.21 &  0.320 &  0.007 &   5.34 &   0.23 &   3.11 &   0.20 &    -  \\
   NGC~2768 &  Group & 36 &  2.251 &  0.014 &   1.82 &   0.21 &  0.285 &  0.006 &   4.65 &   0.20 &   2.84 &   0.18 &  -1.16\\
   NGC~2778 &  Group & 50 &  2.192 &  0.012 &   1.55 &   0.15 &  0.294 &  0.004 &   4.44 &   0.15 &   2.92 &   0.13 &  -0.65\\
   NGC~3377 &  Leo~I & 55 &  2.190 &  0.011 &   1.84 &   0.13 &  0.234 &  0.004 &   4.01 &   0.13 &   2.63 &   0.12 &  -0.27\\
   NGC~3379 &  Leo~I & 59 &  2.323 &  0.010 &   1.46 &   0.12 &  0.287 &  0.004 &   4.75 &   0.12 &   2.76 &   0.11 &  -0.34\\
   NGC~3384 &  Leo~I & 59 &  2.174 &  0.008 &   2.05 &   0.11 &  0.265 &  0.003 &   3.96 &   0.12 &   2.95 &   0.11 &    -  \\
   NGC~3412 &  Leo~I & 52 &  2.005 &  0.012 &   2.23 &   0.14 &  0.201 &  0.004 &   3.43 &   0.14 &   2.74 &   0.13 &    -  \\
   NGC~3489 &  Leo~I & 67 &  2.005 &  0.011 &   3.63 &   0.11 &  0.151 &  0.003 &   2.51 &   0.11 &   2.40 &   0.10 &  -1.62\\
   NGC~3608 &  Group & 57 &  2.264 &  0.011 &   1.59 &   0.13 &  0.288 &  0.004 &   4.43 &   0.13 &   2.76 &   0.12 &  -0.28\\
   NGC~4365 &  Virgo & 49 &  2.387 &  0.012 &   1.72 &   0.14 &  0.303 &  0.004 &   4.96 &   0.15 &   3.11 &   0.13 &    -  \\
   NGC~4374 &  Virgo & 58 &  2.443 &  0.011 &   1.43 &   0.13 &  0.295 &  0.004 &   4.79 &   0.13 &   2.54 &   0.11 &  -0.86\\
   NGC~4382 &  Virgo & 48 &  2.234 &  0.012 &   2.49 &   0.14 &  0.227 &  0.004 &   3.32 &   0.15 &   2.59 &   0.14 &    -  \\
   NGC~4406 &  Virgo & 48 &  2.373 &  0.010 &   1.61 &   0.16 &  0.286 &  0.005 &   4.65 &   0.16 &   2.82 &   0.15 &    -  \\
   NGC~4434 &  Virgo & 36 &  2.080 &  0.016 &   1.77 &   0.19 &  0.250 &  0.006 &   3.90 &   0.20 &   2.30 &   0.18 &    -  \\
   NGC~4458 &  Virgo & 32 &  2.004 &  0.024 &   2.15 &   0.22 &  0.245 &  0.006 &   3.76 &   0.23 &   2.43 &   0.20 &  -0.28\\
   NGC~4464 &  Virgo & 39 &  2.162 &  0.014 &   1.92 &   0.18 &  0.235 &  0.005 &   4.06 &   0.19 &   2.37 &   0.17 &    -  \\
   NGC~4472 &  Virgo & 55 &  2.477 &  0.012 &   1.42 &   0.13 &  0.340 &  0.004 &   5.03 &   0.13 &   3.37 &   0.12 &    -  \\
   NGC~4473 &  Virgo & 78 &  2.264 &  0.007 &   1.42 &   0.09 &  0.300 &  0.003 &   4.62 &   0.09 &   3.10 &   0.08 &    -  \\
   NGC~4552 &  Virgo & 60 &  2.384 &  0.011 &   1.40 &   0.13 &  0.308 &  0.004 &   4.86 &   0.13 &   2.93 &   0.11 &  -0.22\\
   NGC~4564 &  Virgo & 46 &  2.170 &  0.013 &   1.78 &   0.15 &  0.312 &  0.005 &   4.79 &   0.16 &   3.00 &   0.14 &    -  \\
   NGC~4621 &  Virgo & 68 &  2.379 &  0.010 &   1.43 &   0.11 &  0.360 &  0.003 &   5.21 &   0.11 &   3.17 &   0.10 &    -  \\
   NGC~4649 &  Virgo & 47 &  2.524 &  0.013 &   1.61 &   0.16 &  0.325 &  0.005 &   5.00 &   0.16 &   2.81 &   0.14 &  -0.23\\
   NGC~4660 &  Virgo & 64 &  2.290 &  0.009 &   1.35 &   0.11 &  0.319 &  0.003 &   4.72 &   0.11 &   2.70 &   0.10 &    -  \\
   NGC~4673 &   Coma & 43 &  2.356 &  0.013 &   1.93 &   0.16 &  0.269 &  0.004 &   4.14 &   0.16 &   2.81 &   0.15 &  -0.46\\
   NGC~4692 &   Coma & 30 &  2.389 &  0.019 &   1.51 &   0.22 &  0.308 &  0.007 &   4.65 &   0.24 &   3.07 &   0.20 &    -  \\
   NGC~4789 &   Coma & 45 &  2.421 &  0.014 &   1.54 &   0.15 &  0.299 &  0.004 &   4.46 &   0.16 &   2.84 &   0.14 &    -  \\
   NGC~4807 &   Coma & 46 &  2.289 &  0.012 &   1.61 &   0.15 &  0.282 &  0.004 &   4.50 &   0.15 &   3.17 &   0.14 &    -  \\
   NGC~4827 &   Coma & 31 &  2.390 &  0.019 &   1.49 &   0.21 &  0.310 &  0.007 &   4.84 &   0.23 &   2.80 &   0.20 &    -  \\
   NGC~4839 &   Coma & 42 &  2.443 &  0.016 &   1.50 &   0.16 &  0.313 &  0.005 &   5.19 &   0.17 &   3.01 &   0.15 &    -  \\
   NGC~4840 &   Coma & 32 &  2.363 &  0.020 &   1.66 &   0.22 &  0.313 &  0.006 &   5.07 &   0.22 &   2.61 &   0.20 &    -  \\
  NGC~4841A &   Coma & 47 &  2.389 &  0.014 &   1.33 &   0.15 &  0.303 &  0.004 &   4.77 &   0.15 &   2.74 &   0.13 &  -0.20\\
  NGC~4841B &   Coma & 40 &  2.232 &  0.014 &   1.81 &   0.18 &  0.277 &  0.005 &   4.33 &   0.18 &   2.91 &   0.16 &  -0.44\\
   NGC~4860 &   Coma & 41 &  2.436 &  0.014 &   1.53 &   0.16 &  0.330 &  0.005 &   4.98 &   0.17 &   2.81 &   0.15 &    -  \\
   NGC~4864 &   Coma & 43 &  2.309 &  0.014 &   1.85 &   0.17 &  0.278 &  0.005 &   4.51 &   0.17 &   2.79 &   0.15 &  -0.43\\
   NGC~4869 &   Coma & 31 &  2.306 &  0.019 &   1.01 &   0.22 &  0.304 &  0.007 &   4.78 &   0.23 &   2.63 &   0.20 &    -  \\
   NGC~4874 &   Coma & 35 &  2.453 &  0.019 &   1.80 &   0.21 &  0.306 &  0.006 &   4.62 &   0.21 &   2.66 &   0.19 &  -0.28\\
   NGC~4876 &   Coma & 37 &  2.276 &  0.016 &   1.44 &   0.18 &  0.239 &  0.005 &   3.95 &   0.19 &   2.98 &   0.17 &    -  \\
   NGC~4881 &   Coma & 37 &  2.293 &  0.018 &   1.62 &   0.19 &  0.293 &  0.005 &   4.83 &   0.19 &   3.06 &   0.17 &    -  \\
   NGC~4886 &   Coma & 31 &  2.227 &  0.019 &   1.78 &   0.21 &  0.248 &  0.006 &   4.32 &   0.23 &   2.83 &   0.20 &    -  \\
   NGC~4889 &   Coma & 44 &  2.601 &  0.017 &   1.51 &   0.16 &  0.334 &  0.005 &   5.56 &   0.17 &   3.00 &   0.15 &    -  \\
   NGC~4908 &   Coma & 33 &  2.293 &  0.019 &   1.69 &   0.21 &  0.289 &  0.006 &   4.28 &   0.21 &   2.80 &   0.18 &  -0.34\\
   NGC~4926 &   Coma & 44 &  2.432 &  0.014 &   1.56 &   0.16 &  0.315 &  0.005 &   5.23 &   0.16 &   3.06 &   0.14 &    -  \\
   NGC~4952 &   Coma & 45 &  2.453 &  0.014 &   1.61 &   0.15 &  0.266 &  0.005 &   4.41 &   0.16 &   2.93 &   0.14 &    -  \\
   NGC~5582 &      - & 35 &  2.095 &  0.019 &   1.69 &   0.19 &  0.253 &  0.006 &   4.12 &   0.21 &   2.46 &   0.19 &    -  \\
   NGC~5638 &  Group & 48 &  2.231 &  0.013 &   1.48 &   0.17 &  0.321 &  0.005 &   5.17 &   0.18 &   2.89 &   0.16 &    -  \\
   NGC~5812 &      - & 61 &  2.280 &  0.010 &   1.65 &   0.13 &  0.298 &  0.004 &   4.51 &   0.14 &   2.99 &   0.12 &    -  \\
   NGC~5813 &  Group & 56 &  2.377 &  0.012 &   1.64 &   0.15 &  0.285 &  0.004 &   4.45 &   0.15 &   2.85 &   0.13 &  -0.41\\
   NGC~5831 &  Group & 53 &  2.200 &  0.012 &   1.47 &   0.15 &  0.269 &  0.005 &   4.32 &   0.16 &   2.78 &   0.14 &    -  \\
   NGC~5846 &  Group & 48 &  2.344 &  0.013 &   1.52 &   0.18 &  0.307 &  0.005 &   5.05 &   0.18 &   2.55 &   0.16 &  -0.50\\
   NGC~5982 &  Group & 34 &  2.362 &  0.020 &   1.86 &   0.21 &  0.269 &  0.006 &   4.08 &   0.22 &   2.70 &   0.19 &    -  \\
   NGC~6127 &      - & 53 &  2.373 &  0.013 &   1.62 &   0.14 &  0.286 &  0.004 &   4.65 &   0.15 &   2.98 &   0.13 &  -0.22\\
   NGC~6702 &      - & 37 &  2.196 &  0.020 &   2.47 &   0.22 &  0.214 &  0.006 &   3.78 &   0.22 &   2.70 &   0.19 &  -0.30\\
   NGC~6703 &      - & 56 &  2.265 &  0.013 &   2.00 &   0.14 &  0.280 &  0.004 &   4.22 &   0.15 &   2.79 &   0.13 &  -0.54\\
   \hline
   \multicolumn{14}{r}{{\em continued on next page}}\\ \hline
\end{tabular}
\end{table*}
\begin{table*}
\contcaption{}
  \begin{tabular}{llcccccccccccc} \hline 
  Name & env. & S/N & $\log \sigma_0$ & $\epsilon_{\sigma_0}$ & \Hb & $\epsilon_{\rmn{H}\beta}$ & Mg$_2$ &   $\epsilon_{\rmn{Mg}_2}$ & \mgb & $\epsilon_{\rmn{Mg}\/b}$ & \fe & $\epsilon_{<\rmn{Fe}>}$ & \OIIIb\\ \hline
    IC~0613 &  A1016 & 32 &  2.434 &  0.021 &   1.69 &   0.20 &  0.322 &  0.006 &   5.21 &   0.21 &   2.26 &   0.19 &    -  \\
    IC~3973 &   Coma & 38 &  2.341 &  0.016 &   1.78 &   0.17 &  0.283 &  0.005 &   4.46 &   0.19 &   2.51 &   0.17 &    -  \\
    IC~4011 &   Coma & 30 &  1.982 &  0.029 &   1.51 &   0.22 &  0.260 &  0.007 &   4.19 &   0.24 &   2.55 &   0.21 &    -  \\
    IC~4045 &   Coma & 35 &  2.327 &  0.016 &   1.36 &   0.20 &  0.283 &  0.006 &   4.54 &   0.20 &   2.50 &   0.18 &    -  \\
       D024 &   Coma & 35 &  2.279 &  0.015 &   1.54 &   0.21 &  0.287 &  0.006 &   4.38 &   0.21 &   2.70 &   0.19 &    -  \\
       D081 &   Coma & 30 &  2.067 &  0.021 &   1.54 &   0.22 &  0.247 &  0.007 &   4.15 &   0.24 &   2.36 &   0.22 &    -  \\
       D210 &   Coma & 35 &  2.148 &  0.022 &   1.27 &   0.20 &  0.239 &  0.006 &   3.87 &   0.21 &   2.09 &   0.19 &    -  \\
   GMP~1652 &   Coma & 35 &  2.143 &  0.018 &   1.71 &   0.19 &  0.270 &  0.006 &   4.25 &   0.20 &   2.88 &   0.18 &    -  \\
    Z159-43 &   Coma & 31 &  2.415 &  0.021 &   1.50 &   0.23 &  0.315 &  0.007 &   4.84 &   0.23 &   2.86 &   0.20 &  -0.78\\
    Z160-22 &   Coma & 30 &  2.377 &  0.020 &   1.49 &   0.23 &  0.312 &  0.007 &   4.83 &   0.23 &   2.72 &   0.21 &    -  \\
    Z160-23 &   Coma & 30 &  2.218 &  0.018 &   1.45 &   0.23 &  0.283 &  0.007 &   4.73 &   0.24 &   2.98 &   0.21 &    -  \\
    Z160-27 &   Coma & 30 &  2.135 &  0.017 &   1.67 &   0.23 &  0.278 &  0.007 &   4.14 &   0.24 &   2.63 &   0.21 &  -0.30\\
   \hline
\end{tabular}

\medskip

\begin{minipage}{16.5cm}
  Notes: The first column shows the name of the galaxy. If the galaxy
  is found in the NGC or IC (Nilson 1973) catalogues, then it will have
  the name from the corresponding catalogue, in the order of preference
  given above. There are also some galaxies denoted by their name from
  the CGCG (Zwicky et~al. 1961, Zxxx-xx) catalogue, Dressler's (1980,
  Dxxx) catalogue and one galaxy found in the Godwin, Metcalfe \& Peach
  catalogue (1983, GMP~1652). The second column indicates the
  environment. If the galaxy is a member of a well known cluster or
  group, the name is given. For smaller associations we just indicate
  whether the galaxies are in a very low density environment (``-'') or
  belong to a group. The third column shows the effective S/N per~\AA\/
  of the line-strength measurements. The velocity dispersion $\sigma_0$
  and the line-strength indices have been aperture corrected to
  $2r_{norm}=1.19 h^{-1}\,$kpc, equivalent to 3\farcs4 at the distance
  of the Coma cluster. The line-strength indices are calibrated onto
  the Lick/IDS system and corrected to zero velocity dispersion with
  the exception of the \OIIIb\/ index which is calibrated to the \gon\/
  (1993) system.  Mg$_2$ is given in mag, whereas the other
  line-strength indices are given in \AA.
\end{minipage}

\end{table*}

\bsp

\label{lastpage}


\begin{thebibliography}{}
%
%
\bibitem[\protect\citename{Bender, Burstein \& Faber }%
  1993]{ben93} Bender~R., Burstein~D., Faber~S.~M., 1993, ApJ, 411, 153

\bibitem[\protect\citename{Bower, Lucey \& Ellis }%
  1992]{bow92} Bower~R.~G., Lucey~J.~R., Ellis~R.~S., 1992, MNRAS, 254,
  601

\bibitem[\protect\citename{Burstein et~al. }%
  1988]{bur88} Burstein~D., Davies~R.~L., Dressler~A., Faber~S.~M.,
  Lynden-Bell~D., Terlevich~R.~J., Wegner~G., 1988, in Kron~R.~G.  and
  Renzini~A., eds, {\em Towards Understanding Galaxies at Large
    Redshifts}, page~17, Dordrecht, Kluwer

\bibitem[\protect\citename{Butcher \& Oemler }%
  1978]{but78} Butcher H., Oemler A., 1978, ApJ, 219, 18

\bibitem[\protect\citename{Butcher \& Oemler }%
  1984]{but84} Butcher H., Oemler A., 1984, ApJ, 285, 426

\bibitem[\protect\citename{Cole et~al. }%
  2000]{col2000} Cole, S., Lacey, C.\ G., Baugh, C.\ M., Frenk, C.\ S.\ 
  2000, MNRAS, 319, 168

\bibitem[\protect\citename{Colless et~al. }%
  1999]{coll99} Colless~M., Burstein~D., Davies~R.~L., McMahan~R.~K.,
  Saglia~R.~P., Wegner~G., 1999, MNRAS, 303, 813

\bibitem[\protect\citename{Concannon, Rose \& Caldwell }%
  2000]{con2000}Concannon K. D., Rose J. A., Caldwell N., 2000, ApJL,
  536, L19

\bibitem[\protect\citename{Couch et~al. }%
  1998]{cou98} Couch W. J., Barger A. J., Smail I., Ellis R. S.,
  Sharples R. M., 1998, ApJ, 497, 188

\bibitem[\protect\citename{Davies, et~al. }%
  1987]{dav87} Davies R. L., Burstein D., Dressler A., Faber S. M.,
  Lynden-Bell D., Terlevich R. J., Wegner G., 1987, ApJS, 64, 581

\bibitem[\protect\citename{de Jong \& Davies }%
  1997]{dej97} de Jong R. S., Davies R. L., 1997, MNRAS, 285, L1

\bibitem[\protect\citename{Djorgovski \& Davis }%
  1987]{djo87} Djorgovski~S., Davis~M, 1987, ApJ, 313, 59

\bibitem[\protect\citename{Dressler }%
  1980]{dre80} Dressler A., 1980, ApJS, 42, 565

\bibitem[\protect\citename{Dressler }%
  1984]{dre84} Dressler A., 1984, ApJ, 281, 512

\bibitem[\protect\citename{Dressler et~al. }%
  1987]{dre87} Dressler A., Lynden-Bell D., Burstein D., Davies R. L.,
  Faber S. M., Terlevich R. J., Wegner G., 1987, ApJ, 313, 42

\bibitem[\protect\citename{Dressler, et~al. }%
  1997]{dre97} Dressler A. et~al., 1997, ApJ, 490, 577

\bibitem[\protect\citename{Ellis et~al. }%
  1997]{ell97} Ellis R. S., Smail I., Dressler A., Couch W. J., Oemler
  A. J., Butcher H., Sharples R. M., 1997, ApJ, 483, 582

\bibitem[\protect\citename{Ferreras, Charlot \& Silk }%
  1999]{fer99} Ferreras~I., Charlot~S., Silk~J. 1999, ApJ, 521, 81

\bibitem[\protect\citename{Fisher, Franx \& Illingworth }%
  1996]{fis96} Fisher~D., Franx~M., Illingworth~G., 1996, ApJ, 459, 110

\bibitem[\protect\citename{Gonz\'{a}lez }%
  1993]{gon93} Gonz\'{a}lez~J.~J., 1993, PhD thesis, University of
  California

\bibitem[\protect\citename{Godwin, Metcalfe \& Peach}%
  1983]{god83} Godwin, J. G., Metcalfe, N., Peach, J. V., 1983, MNRAS,
  202, 113

\bibitem[\protect\citename{Gorgas, Efstathiou \& Aragon-Salamanca }%
  1990]{gor90} Gorgas~J., Efstathiou~G., Salamanca~A.~A., 1990, MNRAS,
  245, 217

\bibitem[\protect\citename{Greggio }%
  1997]{gre97} Greggio~L., 1997, MNRAS, 285, 151

\bibitem[\protect\citename{Hudson et~al. }%
  1999]{hud99} Hudson M.J., Smith R.J., Lucey J.R., Schlegel D.J.,
  Davies R.L., 1999, ApJ, 512, L79

\bibitem[\protect\citename{J{\o}rgensen, Franx \& Kj{\ae}rgaard }%
  1995]{jor95} J{\o}rgensen~I., Franx~M., Kj{\ae}rgaard~P., 1995,
  MNRAS, 276, 1341

\bibitem[\protect\citename{J{\o}rgensen }%
  1997]{jor97} J{\o}rgensen~I., 1997, MNRAS, 288, 161

\bibitem[\protect\citename{J{\o}rgensen }%
  1999]{jor99} J{\o}rgensen~I., 1999, MNRAS, 306, 607

\bibitem[\protect\citename{Kodama et~al. }%
  1998]{kod98} Kodama T., Arimoto N., Barger A. J., Arag'on-Salamanca
  A., 1998, A\&A, 334, 99

\bibitem[\protect\citename{Kuntschner }%
  1998]{kun98b} Kuntschner~H., 1998, PhD thesis, University of Durham

\bibitem[\protect\citename{Kuntschner \& Davies }%
  1998]{kun98} Kuntschner~H., Davies~R.~L., 1998, MNRAS, 295, L29

\bibitem[\protect\citename{Kuntschner }%
  2000]{kun2000} Kuntschner~H., 2000, MNRAS, 315, 184 

\bibitem[\protect\citename{Larson, Tinsley \& Caldwell }%
  1980]{lar80} Larson~R.~B., Tinsley~B.~M., Caldwell~C.~N., 1980,
  ApJ, 237, 692

\bibitem[\protect\citename{Longhetti et~al. }%
  1998]{lon98} Longhetti~M., Rampazzo~R., Bressan~A., Chiosi~C., 1998,
  A\&AS, 130, 251

\bibitem[\protect\citename{Mehlert }%
  1998]{meh98} Mehlert~D., 1998, PhD thesis, Ludwig - Maximilian -
  Universi\"{a}t, M\"{u}nchen

\bibitem[\protect\citename{Mehlert et~al. }%
  2000]{meh2000} Mehlert D., Saglia R. P., Bender R., Wegner G., 2000,
  A\&AS, 141, 449

\bibitem[\protect\citename{Nilson }%
  1973]{nil73} Nilson P., 1973, Uppsala General Catalogue of Galaxies,
  Uppsala Astron. Obs. Ann., 6 (UGC)

\bibitem[\protect\citename{O'Connell }%
  1976]{oco76} O'Connell~R.~W., 1976, ApJ, 206, 370

\bibitem[\protect\citename{Peletier }%
  1989]{pel89} Peletier~R.~F., 1989, PhD thesis, University of
  Groningen

\bibitem[\protect\citename{Peletier }%
  1999]{pel99} Peletier~R.~F., 1999, in Beckman~J.~E. and
  Mahoney~T.~J., eds, ASP Conf. Ser. Vol~187, The Evolution of Galaxies
  on Cosmological Timescales. Astron. Soc. Pac., San Francisco, p.~231

\bibitem[\protect\citename{Press et~al. }%
  1992]{pre92} Press W. H., Teukolsky S. A., Vetterling W. T., Flannery 
  B. P., 1992, Numerical Recipes -- Second Edition, Cambridge
  University Press, UK 

\bibitem[\protect\citename{Renzini \& Ciotti }%
  1993]{ren93} Renzini A., Ciotti L., 1993, ApJL, 416, L49

\bibitem[\protect\citename{Salaris \& Weiss }%
  1998]{sal98} Salaris~M., Weiss~A., 1998, A\&A, 335, 943

\bibitem[\protect\citename{Sandage \& Visvanathan }%
  1978]{san78} Sandage~A., Visvanathan~N., 1978, ApJ, 223, 707

\bibitem[\protect\citename{Smith et~al. }%
  2000]{smi2000} Smith R.J., Lucey J.R., Hudson M.J., Schlegel D.J.,
  Davies R.L., 2000, MNRAS, 313, 469

\bibitem[\protect\citename{Tantalo, Chiosi \& Bressan }%
  1998]{tan98} Tantalo R., Chiosi C., Bressan A., 1998, A\&A, 333, 419

\bibitem[\protect\citename{Terlevich et~al. }%
  1981]{ter81} Terlevich~R., Davies~R.~L., Faber~S.~M., Burstein~D.,
  1981, MNRAS, 196, 381

\bibitem[\protect\citename{Terlevich \& Forbes }%
  2000]{ter2000} Terlevich~A., Forbes~D. A., submitted to MNRAS

\bibitem[\protect\citename{Trager }%
  1997]{tra97} Trager, S. C., 1997, PhD thesis, University of
  California, CA

\bibitem[\protect\citename{Trager et~al. }%
  1998]{tra98} Trager S. C., Worthey G., Faber S. M., Burstein D.,
  Gonzalez J. J., 1998, ApJS, 116, 1

\bibitem[\protect\citename{Trager et~al. }%
  2000a]{tra2000a} Trager~S.~C., Faber~S.~M., Worthey~G.,
  Gonz{\'a}lez~J.~J., 2000a, AJ, 119, 1645

\bibitem[\protect\citename{Trager et~al. }%
  2000b]{tra2000b} Trager~S.~C., Faber~S.~M., Worthey~G.,
  Gonz{\'a}lez~J.~J., 2000b, AJ, 120, 165 

\bibitem[\protect\citename{Tripicco \& Bell }%
  1995]{tri95} Tripicco~M.~J., Bell~R.~A., 1995, AJ, 110, 3035

\bibitem[\protect\citename{Vazdekis et~al. }%
  2000]{vaz2000} Vazdekis~A., Kuntschner H., Davies R. L., Arimoto N.,
  Nakamura O., Peletier R. F., 2000, submitted to ApJL

\bibitem[\protect\citename{van Dokkum et~al. }%
  1998]{vdo98} van Dokkum P. G., Franx M., Kelson D. D., Illingworth G.
  D., Fisher D., Fabricant D., 1998, ApJ, 500, 714

\bibitem[\protect\citename{Weiss, Peletier \& Matteucci }%
  1995]{wei95} Weiss~A., Peletier~R.~F., Matteucci~F., 1995, A\&A, 296,
  73
\bibitem[\protect\citename{Worthey }%
  1998]{wor98} Worthey~G., 1998, PASP, 110, 888

\bibitem[\protect\citename{Wegner et~al. }%
  1999]{weg99} Wegner~G., Colless~M., Saglia~R.~P., McMahan~R.~K.,
  Davies~R.~L., Burstein~D., Baggley~G., 1999, MNRAS, 305, 259

\bibitem[\protect\citename{Worthey, Faber \& Gonz\'{a}lez }%
  1992]{wor92} Worthey~G., Faber~S.~M., Gonz\'{a}lez~J.~J., 1992, ApJ,
  398, 69

\bibitem[\protect\citename{Worthey }%
  1994]{wor94a} Worthey~G., 1994, ApJS, 95, 107 (W94)

\bibitem[\protect\citename{Worthey \& Ottaviani }%
  1997]{wor97} Worthey~G., Ottaviani~D.~L., 1997, ApJS, 111, 377 

\bibitem[\protect\citename{Zwicky et~al. }%
  1961]{zwi61} Zwicky F., Herzog W., Wild P., Karpowicz M., Kowal C.,
  1961-68, Catalogue of Galaxies and Clusters of Galaxies. California
  institute of Technology, Pasadena (CGCG)

\end{thebibliography}
\end{document}